\documentclass[journal,onecolumn]{IEEEtran}
\usepackage{amssymb}
\usepackage[noadjust]{cite}	%to contract citations
\usepackage[utf8]{inputenc}
\usepackage{textcomp}

\usepackage{graphicx}

\usepackage{amsmath}

\usepackage[version=4]{mhchem}
\usepackage{siunitx}
\usepackage{longtable,tabularx}
\usepackage{algpseudocode,algorithm}
\usepackage{pgfplots}
\usepackage{pgf,tikz}
\usetikzlibrary{calc,arrows}
\usepackage{tikzscale}
\usepackage{wasysym}% For \Circle
\usepackage{acronym}

\usepackage{subcaption}
\usepackage{acronym}

\usepackage{todonotes}
\usepackage[capitalise]{cleveref}

\usepackage{autonum}
\usepackage{colortbl}

\usepackage{amsthm}
\usepackage{gensymb}

\newtheorem{thm}{Theorem}
\newtheorem{defn}{Definition}
\newtheorem{assum}{Assumption}
\newtheorem{rem}{Remark}

\newtheorem{prop}{Proposition}

\crefname{assum}{Assumption}{Assumptions}
\crefname{defn}{Definition}{Definitions}
\crefname{sec}{Section}{Sections}
\crefname{fig}{Fig.}{Figs.}
\crefname{prob}{Problem}{Problems}
\crefname{thm}{Theorem}{Theorems}
\crefname{prop}{Proposition}{Propositions}
\crefname{rem}{Remark}{Remarks}
\newcommand{\R}{\mathbb{R}}

\newcommand{\1}{\mathbf{1}}

\newcommand{\mathcalbf}[1]{\boldsymbol{\mathcal{#1}}}

\newcommand{\satisfies}{\models}
\newcommand{\lalways}{\square}
\newcommand{\leventually}{\lozenge}
\newcommand{\luntil}{\,\mathcal{U}\,}
\newcommand{\lrelease}{\,\mathcal{R}\,}
\newcommand{\lnext}{\Circle}
\newcommand{\true}{\textrm{\it{true}}}
\newcommand{\false}{\textrm{\it{false}}}
\newcommand{\xv}{\mathbf{x}}
\newcommand{\uv}{\mathbf{u}}
\newcommand{\wv}{\mathbf{w}}

\newcommand{\agents}{\mathcal{I}}
\newcommand{\tasks}{\mathcal{J}}
\newcommand{\subagents}{\bar{\agents}}
\newcommand{\subtasks}{\bar{\tasks}}
\newcommand{\decoys}{\mathcal{D}}
\newcommand{\threats}{\mathcal{T}}

\newcommand{\asset}{\alpha}

\newcommand{\StateConstraints}{\mathcal{X}}
\newcommand{\InputConstraints}{\mathcal{U}}
\newcommand{\SpecConstraint}{\mathcal{S}}
\newcommand{\DisturbanceConstraints}{\mathcal{W}}
 \newcommand{\minimise}[1]{\underset{#1}{\operatorname{minimise}}}
\newcommand{\st}{\textrm{subject to}}
\newcommand{\BAop}{\mathcal{B}}
\newcommand{\Bop}{\operatorname{B}}
\newcommand{\LexAop}{\mathcal{S}}

\newcommand{\graph}{\mathcal{G}}

\newcommand{\vertex}{\mathcal{V}}
\newcommand{\edges}{\mathcal{E}}
\newcommand{\weights}{W}
\newcommand{\subedges}{\bar{\edges}}
\newcommand{\subgraph}{\bar{\graph}}

\newcommand{\vass}{\mathcal{P}}

\newcommand{\bott}{\operatorname{b}}
\newcommand{\ebottmm}{\bar{\operatorname{e}}}

\newcommand{\allebott}{\operatorname{E}}

\newcommand{\robOp}{\operatorname{r}}

\newcommand{\dt}{\zeta}
\newcommand{\ds}{\eta}
\newcommand{\dist}{\Delta}
\newcommand{\locCon}{\mathcal{H}}
\newcommand{\bigO}{\mathcal{O}}
\newcommand{\vref}{v^{\textrm{ref}}}

\newcommand{\vmax}{v^{\textrm{max}}}
\newcommand{\SafetyBall}{D}
\newcommand{\TrackingCone}{C}
\newcommand{\ApproxTrackingCone}{\hat{\TrackingCone}}
\newcommand{\aperture}{\theta}
\newcommand{\btrange}{R}
\newcommand{\elconstbase}{e}
\newcommand{\elconst}{\bar{\elconstbase}}
\newcommand{\DopplerShift}{\kappa}
\newcommand{\maxdsvar}{{\DopplerShift}^\textrm{max}}
\newcommand{\FakeAsset}{\bar{\asset}}
\newcommand{\cone}{\textrm{cone}}
\newcommand{\bt}{\textrm{burn}}
\newcommand{\doppler}{\textrm{Doppler}}
\newcommand{\positioning}{\textrm{pos}}
\newcommand{\EndStep}{N^\positioning}
\newcommand{\rot}{h}
\newcommand{\RotAng}{\omega}
\newcommand{\SpeedThreat}{s^z}
\newcommand{\burnthrough}{R}
\newcommand{\polytope}{\mathcal{P}}
\newcommand{\DecoyRange}{D}

\newcommand{\BurnThroughSet}{{\mathcal{Z}}}
\newcommand{\ApproxBurnThroughSet}{\hat{\BurnThroughSet}}
\newcommand{\lowshift}{\mathcal{F}}
\newcommand{\TransFreq}{f^{\textrm{Trans}}}
\newcommand{\aux}{\mathbf{\omega}}

\newcommand{\SafeConstraint}{\mathcal{L}}
\newcommand{\RobSpecConstraint}{\mathcalbf{R}^{\SpecConstraint}}
\newcommand{\ApproxRobSpecConstraint}{\hat{\mathcalbf{R}}^{\SpecConstraint}}
\newcommand{\RobAdmConstraint}{\mathcalbf{R}^{\StateConstraints}}
\newcommand{\RobSafeConstraint}{\mathcalbf{R}^{\SafeConstraint}}

\newcommand{\Tc}{T^{\textrm{c}}}
\newcommand{\Tv}{T^{\textrm{v}}}
\newcommand{\tb}{t^{\textrm{b}}}
\newcommand{\lb}{\upsilon^{\textrm{b}}}

\newcommand{\mumin}{\mu_{\textrm{min}}}
\newcommand{\magents}{\agents_{\textrm{mob}}}
\newcommand{\mtasks}{\tasks_{\textrm{mob}}}

\acrodef{ltl}[LTL]{Linear Temporal Logic}
\acrodef{bap}[BAP]{Bottleneck Assignment Problem}
\acrodef{seqbap}[SeqBAP]{Sequential Bottleneck Assignment Problem}
\acrodef{lexbap}[LexBAP]{Lexicographic Bottleneck Assignment Problem}
\acrodef{mptp}[MTPP]{Minimum-Time Positioning Problem}
\acrodef{milp}[MILP]{Mixed Integer Linear Program}
\acrodef{uav}[UAV]{Unmanned Aerial Vehicle}
\title{
Temporal Logic Planning for Minimum-Time Positioning of Multiple Threat-Seduction Decoys  
}

\author{Tony A.\ Wood$^1$, Mitchell Khoo$^1$, Elad Michael$^1$, Chris Manzie$^1$, and Iman Shames$^2$
\thanks{$^1$Department of Electrical and Electronic Engineering, University of Melbourne, Parkville, VIC, 3010, Australia}
\thanks{$^2$School of Engineering, Australian National University, Acton, ACT, 0200, Australia.}
\thanks{Emails: wood.t@unimelb.edu.au (Tony A.\ Wood), khoom1@student.unimelb.edu.au (Mitchell Khoo), eladm@student.unimelb.edu.au (Elad Michael), manziec@unimelb.edu.au (Chris Manzie), iman.shames@anu.edu.au (Iman Shames).}
}

\begin{document}

\maketitle

\begin{abstract} %should be 100 to 200 words
Reusable decoys offer a cost-effective alternative to the single-use hardware commonly applied to protect surface assets from threats. Such decoys portray fake assets to lure threats away from the true asset. To deceive a threat, a decoy first has to position itself such that it can break the radar lock. Considering multiple simultaneous threats, this paper introduces an approach for controlling multiple decoys to minimise the time required to break the locks of all the threats. The method includes the optimal allocation of one decoy to every threat with an assignment procedure that provides local position constraints to guarantee collision avoidance and thereby decouples the control of the decoys. A crude model of a decoy with uncertainty is considered for motion planning. The task of a decoy reaching a state in which the lock of the assigned threat can be broken is formulated as a temporal logic specification. To this end, the requirements to complete the task are modelled as time-varying set-membership constraints. The temporal and logical combination of the constraints is encoded in a mixed-integer optimisation problem. To demonstrate the results a simulated case study is provided. 
\end{abstract}

% \section*{Nomenclature}
% % \noindent(Nomenclature entries should have the units identified)
% {\renewcommand\arraystretch{1.0}
% \noindent\begin{longtable*}{@{}l @{\quad=\quad} l@{}}
% $\asset$ & asset\\
% $y$ & asset position [m]\\
% $\threats$ & set of threats\\
% $m$ & number of threats\\
% $z_j$ & position of threat j [m]\\
% $\decoys$ & set of decoys\\
% $n$ & number of decoys\\
% $p_i$ & positon of decoy i\\
% $t$ & time [s]\\
% $\aperture$  & half of the aperture of threat tracting cone [rad] \\
% $\elconst_j$ & electromagnet constant for decoy $i$ jamming threat $j$ [$\sqrt{\textrm{m}}$]\\
% $\vmax$ & maximum decoy velocity component [m/s]\\
% $\vref$ & reference decoy component-wise velocity for variation of local bounds [m/s]\\
% $\umax$ & maximal commanded velocity component by planning controller [m/s]\\
% $A_k$ & saturation value of local position bound parameter for bottleneck order $k$\\
% $\DopplerShift$ & Doppler shift [1/s]\\
% $\maxdsvar$ & Maximum toleratedl Doppler shift [1/s]\\
% \multicolumn{2}{@{}l}{Subscripts}\\
% $i$ & associated to decoy $i$\\
% \end{longtable*}}

% {\color{red}
\section{Introduction}
In this paper, we consider a method to protect a surface asset from multiple incoming ballistic threats. Upon detection of a threat, there is typically only a limited time, in the order of 100 seconds, until the threat reaches the asset. Therefore, a fast real-time defence response is required. We assume the availability of a group of so-called threat-seduction decoys equipped with electronic countermeasures \cite{Adamy2000Book}. Common decoy-based responses to this type of threat scenario rely on single-use hardware, e.g. products by Terma, Chemring Countermeasures, and Rheinmetall.  %\cite{TermaSKWS,Chemring2011,Rheinmetall2010}
To reduce operation costs we consider decoys that consist of reusable \acp{uav} as proposed in~\cite{Shames2017CDC}. The objective of such a decoy is to lure a threat away from the asset by portraying a fake asset. To this end, the decoy has to position itself such that the target lock of threat is broken and the radar signature of the fake asset resembles the signal of the true asset. This process is referred to as spoofing or deceptive jamming~\cite{Adamy2000Book}. In a scenario with multiple threats, coordinating and controlling the group of decoys involves the problems of task assignment, collision avoidance, and motion control. While reusable \acp{uav} provide the potential of significant cost savings, they can typically only travel at lower speeds in comparison to single-use decoys and therefore amplify the time-criticality of the motion control problem. An overview of \ac{uav} control methods and challenges in the context of defence against threats can be found in~\cite{Shima2009Book}. In~\cite{Jang2019JoAE} the objective of controlling a fleet of drones to collaboratively disrupt multiple stationary radars is considered. In contrast, the objective we consider here involves jamming the radar systems of threats that are travelling at high speeds. 
 
\emph{Task assignment} refers to the problem of assigning a set of tasks to a set of agents. Assignment problems can be expressed as optimisation problems with permutations of integers as decision variables with a wide range of different possible objectives~\cite{Burkard2012Book,Pentico2007EJoOR}. The consideration of task assignment for \ac{uav}s has been studied extensively, e.g., in~\cite{Bertuccelli2009GNaCC,Bethke2008RaAM}. In \cite{Edison2011CaOR} the problem of assigning tasks to a group of \ac{uav}s, each with the capability of fulfilling multiple tasks in sequence, is considered. Those tasks consist of predefined poses that need to reached and the objective is to minimise the combined distances the \ac{uav}s travel. To simultaneously positioning multiple threat-seduction decoys, we consider the objective of minimising the largest positioning time among all assigned decoys. Finding an assignment, where the largest individual cost of pairing an agent to a task is minimised, is referred to as a \ac{bap}. An assignment where not only the bottleneck cost but also the second, third, and all consecutive pairing costs are to be hierarchically minimised is called a \ac{lexbap}.

In \cite{Morgan2016IJoRR} \ac{uav}s are assigned to relative positions within a swarm and collision avoidance is achieved by convexifying inter-agent distance constraints. Many different tailored strategies exist for avoiding collisions among multiple mobile agents  that are assigned to interchangeable target destinations. In \cite{Turpin2014AR,Gravell2021CEP} collisions are avoided by introducing delayed starting times for pre-computed trajectories to assigned destinations. In \cite{Bertrand2014AJ,Wu2019ISMRMAS} tasks are assigned based on optimisation problems that incentivise agents to avoid situations that lead to collisions. Other approaches exploit properties of optimal assignments to obtain collision-free, straight-line trajectories for agents modelled as point masses~\cite{Shames2011IJoRaNC,MacAlpine2015CoAI}. A minimum spacing of agent initial positions and destinations, that guarantees collision avoidance for agents with finite extent following straight-line trajectories towards the destinations that minimise the sum of the travelled distances, is derived in \cite{Turpin2014IJRR}. To provide collision avoidance guarantees for decoys with a known finite extent in the case where the destinations are assigned to minimised the bottleneck distance, we use the method introduced in \cite{Wood2020RAL} where time-varying convex position constraints are derived from the optimal solution of a \ac{lexbap}. Given that the optimal assignment is sufficiently robust to variations of the distances between the initial and final positions of the agents, the considered local position constraints also allow for trajectories that deviate from straight lines and constant speeds. 

For motion control with more complex objectives than point-to-point guidance formal methods, such as \ac{ltl}, are becoming increasingly popular for planning the trajectories of mobile agents~\cite{Tabuada2005TAC,Belta2007RaAM,Kress2009ToR,Wongpiromsarn2012TAC}. \ac{ltl} enables the formulation of tasks with logical statements and timing constraints by combining propositional logic with temporal operators. In~\cite{Karaman2008CDC, Wolff2014ICRA, Gol2015Automatica, Frick2017LCSS, Sessa2018HSCC} optimal control synthesis problems for discrete-time dynamical systems are formulated as mixed-integer optimisation problems. Control of systems with temporal logic specification in the presence of uncertainty is addressed with robust formulations in~\cite{Wolff2012CDC, Bloem2012JCSS, Frick2017LCSS, Sessa2018HSCC} and with stochastic approaches in~\cite{Ding2011IFAC, Lahijanian2011ToR, Kamgarpour2017Automatica}. We follow the approaches used in \cite{Frick2017LCSS,Sessa2018HSCC} where atomic propositions are defined such that they relate to the conditions of the state of a dynamical system lying in particular polyhedral sets.

The main contribution of this paper is modelling the positioning task of a threat-seduction decoy as a temporal logic specification satisfaction problem where time-varying set-membership constraints on the velocity and position are defined as propositions. To coordinate between multiple decoys that are defending against multiple threats in parallel, we apply the sequential bottleneck assignment procedure introduced in~\cite{Wood2020RAL}. This involves assigning threats to available decoys and computing time-varying safe sets for every decoy to avoid collisions. To this end, we derive static target jamming locations and use the distance of the initial decoy positions to them as assignment weights. We prove that the derived target locations are the positions that remain suitable for jamming for the longest possible time. The local safe sets, obtained from assignment robustness, decouple the motion planning of the individual decoys. We therefore model the conditions required for an individual decoy to break the lock of its assigned threat as polyhedral constraints on its state and formulate the positioning task as an \ac{ltl} formula. We also derive a method to encode the completion time of the positioning task. We apply the robust control synthesis methods introduced in~\cite{Frick2017LCSS} to obtain a commanded input sequence that minimises the positioning time by solving a mixed-integer optimisation problem that is computationally tractable. We demonstrate that the added flexibility, provided by having time-varying set-valued targets and letting the timing of the satisfaction of these target conditions be part of the optimisation problem, reduces the positioning time in comparison to the strategy of simply navigating decoys to predefined static targets.

The rest of this article is structured as follows. Background theory and definitions are provided in \cref{sec:preliminaries}. We introduce the threat scenario and the minimal-time positioning problem in~\cref{sec:prob}. In \cref{sec:TaskAssignment} we derive suitable target jamming locations that are used to estimate the positioning times for every possible decoy-threat pair, make an assignment of threats to decoys, and provide local safe sets for every decoy. In~\cref{sec:specification} we model a time-varying polyhedral set of decoy states for each subtask required for jamming a threat. An \ac{ltl} specification that combines these subtasks and captures the positioning objective of a decoy is then derived. In~\cref{sec:RobOpt} we encode the specification into a mixed-integer optimisation problem that minimises the individual positioning time. The derived concepts are applied and evaluated in simulation in~\cref{sec:CaseStudy} before concluding remarks are presented in~\cref{sec:conclusion}.

\section{Preliminaries}\label{sec:preliminaries}
The smallest integer greater than or equal to scalar $b$ is denoted by $\lceil b\rceil$.
We use $[A]_q$ to denote the $q$-th row of matrix $A$; if $a$ is vector, $[a]_q$ denotes the $q$-the element of $a$. Furthermore, $\|a\|_2:=\sqrt{\sum_{q=1}^{n_a}[a]_q^2}$ denotes the Euclidean norm and $\|a\|_\infty:=\max_{q \in \{1,\dots,n_a\}}\left|[a]_q\right|$ denotes the infinity norm of vector $a \in \R^{n_a}$. Throughout this paper, bounds on vectors are considered with respect to the infinity norm which leads to affine constraints. These bounds can be obtained from other norms with an appropriate scaling, i.e., $\|a\|_2 \leq b$ can be translated to $\|a\|_\infty \leq \sqrt{n_a}b$ for $a \in \R^{n_a}$. 
 
In~\cref{sec:pre_assignment} we define the tools required for the assignment procedure introduced in~\cite{Wood2020RAL} and in~\cref{sec:pre_CollAvoid} restate its properties that provide collision avoidance guarantees. Then, in \cref{sec:pre_LTL}, as a framework to capture complex motion planning constraints, we provide definitions for a specific class of \ac{ltl} formula considered in~\cite{Frick2017LCSS}.

\subsection{Task Assignment}\label{sec:pre_assignment}
Given a set of $m$ agents, $\agents=\{i_1,\dots,i_m\}$, and a set of $n$ tasks, $\tasks=\{j_1,\dots,j_n\}$, with $m \geq n$, we define the complete bipartite \emph{assignment graph} $\graph:=(\agents,\tasks,\edges)$, with vertex set, $\vertex:=\agents \cup \tasks$, and edge set, $\edges:=\agents \times \tasks$. 
\begin{defn}[Assignment]
Let $\Pi:=\{\Pi_{i,j}\}_{(i,j)\in\edges}$ be a set of binary variables where $\Pi_{i,j}=1$ corresponds to task $j$ being assigned to agent $i$. $\Pi$ is an \emph{assignment} of a subset of tasks, $\subtasks \subseteq \tasks$, to a subset of agents, $\subagents \subseteq \agents$, with respect to the subset of edges, $\hat{\edges} \subseteq \subagents \times \subtasks$, if all tasks in $\subtasks$, are assigned to one agent in $\subagents$ and all these agents are assigned to at most one task. The set of such assignments for subgraph $\hat{\graph}:=(\subagents,\subtasks,\hat{\edges})$ is $
\vass_{\subagents,\subtasks}(\hat{\edges}):= \big\{\{\Pi_{i,j}\}_{(i,j)\in\edges}\,\big|\, 
\forall (i,j) \in \edges : \Pi_{i,j} \in \{0,1\},
\forall j \in \subtasks : \sum_{i \in \{i'\in\subagents|(i',j)\in\hat{\edges}\}}\Pi_{i,j} = 1, 
\forall i \in \subagents : \sum_{j \in \{j'\in\subtasks|(i,j')\in\hat{\edges}\}}\Pi_{i,j} \leq 1 
\big\}
$. 
\end{defn} 

To evaluate an assignment, we assume there is a set of \emph{assignment weights}, $\weights:=\{W_{i,j}\}_{(i,j)\in\edges}$, where $W_{i,j} \geq 0$ is the cost of assigning tasks $j \in \tasks$ to agent $i \in \agents$.
\begin{defn}[Bottleneck assignment]\label{defn:bottass}
Consider a subgraph of the assignment graph, $\hat{\graph}=(\subagents,\subtasks,\hat{\edges})$, with $\subagents\subseteq\agents$, $\subtasks\subseteq\tasks$, and $\hat{\edges}\subseteq\subagents\times\subtasks$. The set of \emph{bottleneck minimising assignments} is $\BAop_{\subagents,\subtasks}(\hat{\edges},\weights):= \arg\min_{\Pi\in \vass_{\subagents,\subtasks}(\hat{\edges})} \bott(\Pi,\hat{\edges},\weights)$ and $\Bop_{\subagents,\subtasks}(\hat{\edges},\weights):= \min_{\Pi\in \vass_{\subagents,\subtasks}(\hat{\edges})} \bott(\Pi,\hat{\edges},\weights)$ 
is the so-called \emph{bottleneck weight}, where  
$
  \bott(\Pi,\hat{\edges},\weights) := \max_{(i,j)\in\hat{\edges}} \:\Pi_{i,j}W_{i,j} 
$
is the largest weight corresponding to any assigned agent-task pair in $\hat{\edges}$ for a given assignment, $\Pi \in \vass_{\subagents,\subtasks}(\hat{\edges})$. The set of edges in $\hat{\edges}$ with weight equal to the bottleneck is 
$
\allebott_{\subagents,\subtasks}(\hat{\edges},\weights):=\big\{(i,j)\in\hat{\edges}\,\big|\,W_{i,j}=\Bop_{\subagents,\subtasks}(\hat{\edges},\weights)\big\}
$.
\end{defn}

To quantify the difference in cost of a bottleneck optimal assignment to the next best assignment when a particular edge is discarded, we consider the notion of robustness margin defined in \cite{Wood2020RAL} that is related to the concept of allowable perturbation introduced in~\cite{Michael2021arXiv}.
\begin{defn}[Robustness margin]\label{def:criticaledge}
  Given a subset of agents, $\subagents\in\agents$, and a subset of tasks, $\subtasks\in\tasks$,  consider the complete bipartite subgraph of the assignment graph,
$\subgraph:=(\subagents,\subtasks,\subedges)$, with $\subedges:=\subagents\times\subtasks$. For $|\subedges|>1$, the set $\ebottmm_{\subagents,\subtasks}(\weights):= \arg\max_{(i,j)\in\allebott_{\subagents,\subtasks}(\subedges,\weights)}\Bop_{\subagents,\subtasks}(\subedges\setminus\{(i,j)\},\weights)$ contains all so-called \emph{maximum-margin bottleneck edges} and $\robOp_{\subagents,\subtasks}(\weights):= \max_{(i,j)\in\allebott_{\subagents,\subtasks}(\subedges,\weights)}\Bop_{\subagents,\subtasks}(\subedges\setminus\{(i,j)\},\weights) - \Bop_{\subagents,\subtasks}(\subedges,\weights)$ is the corresponding \emph{robustness margin}. For $|\subagents|=|\subtasks|=|\subedges|=1$, the maximum-margin bottleneck edge is set to be the singleton edge $\ebottmm_{\subagents,\subtasks}(\weights)=\subedges$ and the robustness margin is assumed to be infinity, $\robOp_{\subagents,\subtasks}(\weights)=\infty$.
\end{defn}

The definition of bottleneck minimising assignments does not fully determine all agent-task pairings. We therefore consider the sequence of assignment objectives with a decreasing hierarchy defined in \cite{Wood2020RAL}.
\begin{defn}[Sequential bottleneck assignment]\label{defn:lexass} 
An assignment, $\Pi^*$, is sequential bottleneck optimising if it is bottleneck minimising for the assignment graph, $\graph=(\agents,\tasks,\edges)$, and the sequence of subgraphs $\subgraph_2,\subgraph_3,\dots,\subgraph_n$, where $\subgraph_l=(\subagents_l,\subtasks_l,\subedges_l)$ is the complete bipartite graph of the subset of agents, $\subagents_l\subset\agents$, and the subset of tasks, $\subtasks_l\subset\tasks$, obtained by removing a maximum-margin bottleneck agent and tasks from $\subgraph_{l-1}$, i.e.,
$  \Pi^* \in \LexAop_{\agents,\tasks}(\weights) := \big\{\Pi\in\vass_{\agents,\tasks}(\edges)\,\big|\,\forall l\in\{1,\dots,n\} : \Pi\in\BAop_{\subagents_l,\subtasks_l}(\subedges_l,\weights)\big\}$,
where $\subedges_l=\subagents_l\times\subtasks_l$, $\subagents_1=\agents$, $\subtasks_1=\tasks$, 
$    \subagents_l = \subagents_{l-1}\setminus\{i^*_{l-1}\}, \subtasks_l = \subtasks_{l-1}\setminus\{j^*_{l-1}\}$,
with so-called \emph{$l$-th order bottleneck edge},
$ (i^*_l,j^*_l)\in\ebottmm_{\subagents_l,\subtasks_l}(\weights)$, 
and \emph{$l$-th order robustness margin},
$ \mu_l = \robOp_{\subagents_l,\subtasks_l}(\weights)$.
\end{defn}
A sequential bottleneck optimising assignment is unique and lexicographically optimal if the robustness margins of all orders are strictly positive. 

\subsection{Collision Avoidance via Robust Task Assignment}\label{sec:pre_CollAvoid}
Given $m$ mobile agents, $\magents$, with initial centroid positions $p_i(0) \in \R^{n_{\textrm{mob}}}$, $i\in\magents$, we define a set of tasks, $\mtasks$, corresponding to $n \leq m$ target destinations, $g_j\in\R^{n_{\textrm{mob}}}$, $j\in\mtasks$, and consider the distances between agent initial positions and target destinations as assignment weights, $W_{i,j}=\|g_j-p_i(0)\|$, where $\|\cdot\|$ is an arbitrary norm.
\begin{defn}[Local position constraints]\label{defn:LocalConstraint}
  Let tasks $\mtasks$ be allocated to agents $\magents$ by a sequential bottleneck optimising assignment, $\Pi^*=\LexAop_{\magents,\mtasks}(W)$, with $W = \{\|g_j-p_i(0)\|\}_{(i,j)\in\magents\times\mtasks}$. Consider a desired \emph{safety distance} $s$ that is smaller than the smallest robustness margin, $s < \mumin:= \min_{l \in {1,\dots,n}}\mu_l$.
  The local position constraints for an $l$-th order bottleneck agent, $i^*_l$, $l\in\{1,\dots,n\}$, is
  $
      \locCon_{i^*_l}(t,s):=\big\{p\in\R^3\,\big|\,\|p-p_{i^*_l}(0)\| \leq \ds_{i^*_l}(t,s) - \epsilon, \,\|g_{j^*_l}-p\| \leq \dt_{i^*_l}(t,s) - \epsilon \big\}
  $,
  with 
  \begin{subequations}\label{eq:localBounds}
  \begin{align+}
  \ds_{i^*_l}(t,s) &:=
  \begin{cases}
    \ds(t)  &\textrm{if } \ds(t) \leq A_{i^*_l}(s) \,,
    \\ A_{i^*_l}(s) & \textrm{otherwise}\,,
  \end{cases} 
  \\ \dt_{i^*_l}(t,s) &:= A_{i^*_l}(s) + \frac{1}{2}(\mumin - s)- \ds_{i^*_l}(t,s)\,, 
  \\ A_{i^*_l}(s)&:= \min_{l'\in\{1,\dots,l\}} w_{i^*_{l'},j^*_{l'}} + \mu_{l'} - \frac{1}{2}(\mumin + s)\,,
\end{align+} 
\end{subequations}
where $\epsilon>0$ is arbitrarily small and $\ds(t)>0$ is a shared time-varying coordination variable. For any unassigned agent, $i'\in\magents\setminus \{i^*_1,\dots,i^*_n\}$, the local position constraint set is
$
      \locCon_{i'}(t,s):=\big\{p\in\R^3\,\big|\,\|p-p_{i'}(0)\| \leq \ds_{i^*_n}(t,s) - \epsilon \big\}
$.
\end{defn}

The local position constraints ensure that agents en route to their assigned destinations do not collide. 
% \begin{prop}[Proof in \cite{Wood2020RAL}]\label{prop:CollisionAvoidance}
%   Assume tasks $\mtasks$ are allocated to agents $\magents$ with a sequential bottleneck optimising assignment, $\Pi^*=\LexAop_{\magents,\mtasks}(\weights)$, with $W_{i,j}=\|g_j-p_i(t_0)\|$. Let $p_i(t)$ denote the centroid position of agent $i \in \magents$ at time $t$. The set of assigned agents is denoted by $\{i^*_1,\dots,i^*_n\}$, where $i^*_l$ denotes the $l$-th order bottleneck agent, $j^*_l$ denotes the $l$-th order bottleneck task, and $\mu_l$ is the $l$-th order robustness margin. 
%   Assume that the minimal robustness margin is greater than a defined safety distance $s>0$, i.e., $\min_{l \in {1,\dots,n}}\mu_l>s$.
%   At time $t \geq 0$, the distance between the position of any assigned agent, $i^*_l$, $l\in\{1,\dots,n\}$, and any other agent, $i'\in \magents\setminus\{i^*_l\}$, is greater than the safety distance, i.e., $\|p_{i^*_l}(t)-p_{i'}(t)\|>s$, if all agents satisfy their local position constraints, i.e., $p_i(t)\in\locCon_i(t,s)$ for all $i\in\magents$.
% \end{prop}
\begin{prop}[Proof in \cite{Wood2020RAL}]\label{prop:CollisionAvoidance}
   Given a sequential bottleneck optimising assignment, $\Pi^*=\LexAop_{\magents,\mtasks}(\weights)$, with $W = \{\|g_j-p_i(0)\|\}_{(i,j)\in\magents\times\mtasks}$, where $g_j$ is the destination corresponding to task $j \in \mtasks$ and $p_i(t)$ is the centroid position of agent $i \in \magents$ at time $t$. The set of assigned agents is denoted by $\{i^*_1,\dots,i^*_n\}$, where $i^*_l$ denotes the $l$-th order bottleneck agent, $j^*_l$ denotes the $l$-th order bottleneck task, and $\mu_l$ is the $l$-th order robustness margin. 
   Assume that the minimal robustness margin is greater than a defined safety distance $s>0$, i.e., $\min_{l \in {1,\dots,n}}\mu_l>s$.
   At time $t \geq 0$, the distance between the centroid positions of any assigned agent, $i^*_l$, $l\in\{1,\dots,n\}$, and any other agent, $i'\in \magents\setminus\{i^*_l\}$, is greater than the safety distance, i.e., $\|p_{i^*_l}(t)-p_{i'}(t)\|>s$, if all agents satisfy their local position constraints, i.e., $p_i(t)\in\locCon_i(t,s)$ for all $i\in\magents$.
\end{prop}

\subsection{Linear Temporal Logic}\label{sec:pre_LTL}
We use $\mathbf{a}^{L}[k]:=(a[k],a[k+1],\dots,a[k+L])$ to denote a sequence of length $L + 1$ starting at time step $k$, where $a[l]$ is the value of a variable at discrete time step $l$. For a system with state vector, $x[k] \in \R^{n_x}$, at time steps $k$, we consider bounded \ac{ltl} formulae, as in~\cite{Frick2017LCSS}, in positive normal form, see~\cite{Baier2008book}, and without loops, see~\cite{Birkedal2006Arix}. An \ac{ltl} formula combines atomic propositions, which are either satisfied or not satisfied, with propositional logic operators, $\lnot$~(not), $\land$~(and), $\lor$~(or), and temporal operators, $\lnext$~(next), $\luntil$~(until), $\lrelease$~(release), see~\cite{Wolff2014ICRA}. 
\begin{defn}[Atomic propositions]\label{defn:ap}
  An atomic proposition, ${ap}$, is associated with a time-varying polyhedral subset of the state space, $\polytope_{ap}[k] = \big\{ x\in\R^{n_x} \,\big|\, \Phi_{{ap}}[k] x \leq \Psi_{{ap}}[k] \big\}$. A sequence of states of length $L+1$ starting at time step $k$, satisfies atomic proposition ${ap}$  if and only if the state at time step $k$ lies in the associated polyhedral set, $x[k] \in \polytope_{{ap}}[k]$. Let $\xv^{L}[k] \satisfies {ap}$ denote the satisfaction of ${ap}$ by $\xv^{L}[k]$.
\end{defn}
\begin{defn}[\ac{ltl} formula]\label{defn:ltl}
The satisfaction of a formula, $\varphi$, by a sequence of states of length $L+1$ starting at time step $k$ is denoted by $\xv^{L}[k]\satisfies \varphi$; the sequence not satisfying the formula is denoted by $\xv^{L}[k] \not\satisfies \varphi$. The most basic formulas are defined as single atomic propositions, e.g., $\varphi = {ap}$. The propositional logic is defined inductively as
  \begin{align}
      \xv^{L}[k]&\satisfies \lnot\varphi &\Leftrightarrow& && \xv^{L}[k] \not\satisfies\varphi\,,\\
      \xv^{L}[k] & \satisfies \varphi^1 \land \varphi^2 &\Leftrightarrow& && \xv^{L}[k] \satisfies \varphi^1\textrm{ and }\xv^{L}[k] \satisfies \varphi^2\,,\\
      \xv^{L}[k]&\satisfies \varphi^1 \lor \varphi^2 &\Leftrightarrow& && \xv^{L}[k] \satisfies \varphi^1\textrm{ or }\xv^{L}[k] \satisfies \varphi^2\,.
  \end{align}
  The temporal operators are defined as
  \begin{align}
      \xv^{L}[k]&\satisfies \lnext \varphi &\Leftrightarrow& && \xv^{L-1}[k+1] \satisfies \varphi\,,\\
      \xv^{L}[k]&\satisfies \varphi^1 \luntil \varphi^2 &\Leftrightarrow& && \exists l\in\{0,1,\dots, L\} \textrm{ s.t. } \xv^{L-l}[k+l] \satisfies \varphi^2 
      \textrm{ and } \forall l'\in \{0,1,\dots, l-1\} : \xv^{L-l'}[k+l'] \satisfies \varphi^1 \,,\\
      \xv^{L}[k]& \satisfies \varphi^1 \lrelease \varphi^2 &\Leftrightarrow& && \forall l\in \{0,1,\dots, L\} : \xv^{L-l}[k+l] \satisfies \varphi^2 
      \textrm{ or }\exists l'\in\{0,1,\dots, l-1\} \textrm{ s.t. } \xv^{L-l'}[k+l'] \satisfies \varphi^1\,.
  \end{align}
\end{defn}
 As shorthand we define the additional temporal operators $\leventually\varphi:=\true\luntil\varphi$ (eventually) and $\lalways\varphi := \false \lrelease \varphi$ (always), where $\xv^{L}[k]\satisfies\true$ and $\xv^{L}[k]\not\satisfies\false$ for all $\xv^L[k]$.

\section{Problem Formulation}
\label{sec:prob}
Given a surface asset, $\asset$, we consider $n$ incoming threats, $\threats:=\{\tau_1,\dots,\tau_n\}$, and $m\geq n$ controllable decoys, $\decoys:=\{\delta_1,\dots,\delta_m\}$. The asset, $\asset$, is located at position~$y(t)\in\{y\in\R^3\,\big|\,[y]_3=0\}$ and is assumed to move with constant velocity, $\dot{y}(t) = v^\asset$, for time $t\geq t_0$, where $t_0 = 0$ is the time the initial planning decision is made. The position of threat~$j \in \threats$ is denoted by $z_j(t)\in\R^3$. For the considered phase of the threat trajectory, we model the velocity by
\begin{align}\label{eq:ThreatVelocity_pre}
    \dot{z}_j(t) = \frac{\SpeedThreat}{\|y(t)-z_j(t)\|_2}\left(y(t)-z_j(t)\right)\,,
\end{align}
where $\SpeedThreat\gg \|v^\asset\|_2$ is a constant speed parameter. A decoy, $i\in\decoys$, is represented by its centroid position, $p_i(t)\in\R^3$, and velocity, $v_i(t)=\dot{p}_i(t)$. We assume that the entire body of decoy~$i$ lies within the set, 
$
    \SafetyBall_i(t):=\big\{b\in\R^3\,\big|\, 2 \|b-p_i(t)\|_\infty\leq d \big\}
$. 

\subsection{Threat-Seduction Strategy}
\begin{figure}%[thb]
    \centering
    \includegraphics[width=0.98\linewidth]{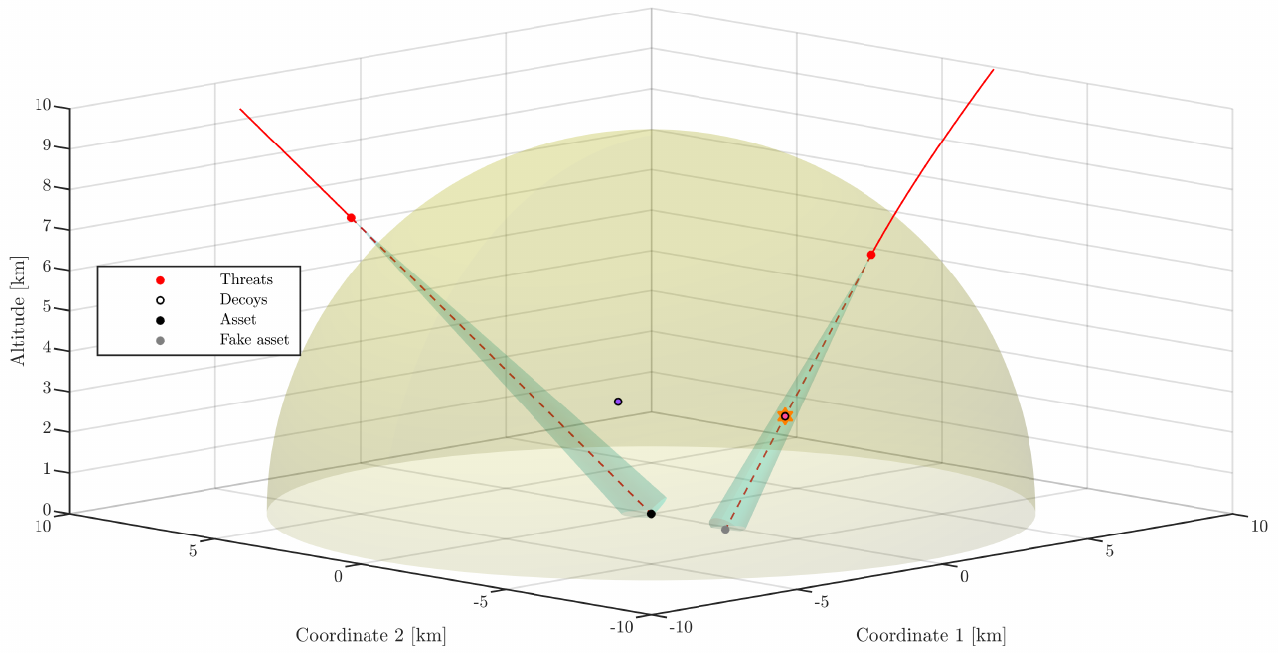}
    \caption{Threat scenario and seduction strategy: threat tracking cones illustrated by cyan surfaces with dashed red axis; for the left threat the burn-through range with respect to the purple decoy is shown with yellow surface; the right threat is being seduced by the pink decoy.} 
    \label{fig:ThreatScenario}
\end{figure}

The defence strategy considered here and first introduced in \cite{Shames2017CDC} consists of controlling the decoys such that each threat is lured away from the asset by one decoy. For decoy $i\in\decoys$ to deceive threat $j\in\threats$, it first has to reach a state in which it can break the target lock of the threat. In the following step, the decoy then seduces the threat away from the asset by making it follow a fake asset, $\FakeAsset_i$. 
The fake asset is portrayed to be positioned at $\bar{y}_i(t)$ which is the point where the line through the threat position, $z_j(t)$, and the decoy position, $p_i(t)$, intersects with the ground surface, see \cref{fig:ThreatScenario}. During the seduction phase the threat velocity is oriented towards the fake asset, i.e., $y(t)$ is replaced with $\bar{y}_i(t)$ in \eqref{eq:ThreatVelocity_pre}.

To be able to interfere with the navigation of threat $j$, a decoy has to position itself inside the time-varying set of locations that can be tracked by the threat, 
  \begin{align}\label{eq:trackingcone}
    \TrackingCone_j(t):=\big\{p\in\R^3\,\big|\, 
    \|p-z_j(t)\|_2 \SpeedThreat\cos(\aperture) \leq (p-z_j(t))^\top \dot{z}_j(t),\, \|p-z_j(t)\|_2 \leq \|y(t)-z_j(t)\|_2 
      \big\}\,,
  \end{align}
referred to as the \emph{tracking cone}. The tracking cone consists of all positions that lie closer to the threat than the asset does and within a pointed cone with apex at the treat position, $z_j(t)$, axis aligned with the threat velocity, $\dot{z}_j(t)$,  and aperture $2\aperture\in\left(0,\frac{\pi}{2}\right)$, as shown in \cref{fig:TrackingCone_pure}. Once the decoy lies in the traction cone, $p_i(t)\in\TrackingCone_j(t)$, it has to generate a radar return that mimics the signal of the asset to break the lock of the threat. In~\cite{Adamy2000Book} a necessary condition for which this is possible is introduced. The so-called \emph{burn-through range}, $\btrange_{i,j}(t)$, is the minimum distances between the asset and threat~$j$ for which jamming with decoy~$i$ is effective, i.e., the threat cannot be made to lock onto the decoy if 
$
    \|y(t)-z_j(t)\|_2 < \btrange_{i,j}(t)
$.
The magnitude of the burn-through range is a function of the distance of the decoy from the threat and is given by
\begin{align}\label{eq:BurnThroughRange}
    \btrange_{i,j}(t):=\elconst_j \sqrt{\|p_i(t)-z_j(t)\|_2}
    \,,
\end{align} 
where $\elconst_j$ is a constant that depends on the antenna properties of the threat radar. We refer to \cref{fig:ThreatScenario} for an illustration of the burn-through range and to~\cite{Adamy2000Book, Kaptan2012Thesis} for its derivation. 

If decoy~$i$ reaches the tracking cone, $\TrackingCone_j(t)$, while threat~$j$ lies outside of the burn-through range, $\btrange_{i,j}$, then the seduction of the threat can be initialised by the decoy portraying the fake asset, $\FakeAsset_{i}$. For successful seduction, the fake asset has to evolve in a way the threat cannot detect being different from the true asset. In particular, the decoy has to manoeuvre such that the Doppler shift it causes is similar to the one produced by the asset,
$
      |\DopplerShift_{i,j}(t)-\DopplerShift_{\asset,j}(t)|\leq\maxdsvar 
$,
where $\maxdsvar>0$ is the tolerance in frequency that will not be detected. The Doppler shift resulting from transmission source~$\sigma\in\decoys\cup\asset$ and observed by the threat~$j$ is given by
\begin{align}\label{eq:ds}
    \DopplerShift_{\sigma,j}(t)= \frac{\TransFreq}{\SpeedThreat c}\left(v(t)-\dot{z}_j(t)\right)^\top\dot{z}_j(t)\,,
\end{align} with $v(t)=v^\asset$ if the threat is locked onto the asset, i.e., $\sigma=\asset$, and $v(t)=v_i(t)$ if the threat is locked onto the fake asset, i.e., $\sigma=i$, where $\TransFreq$ is the transmission frequency, and $c$ is the speed of light. 

\subsection{Minimising Longest Positioning Time}
The overriding goal in controlling the decoys is to seduce all threats away from the asset as reliably as possible. The quicker a decoy can position itself such that it can start the process of deceiving a threat, the better the chances are for successfully luring the threat away from the asset. We therefore aim to control the decoys such that the time it takes to start the seduction process of all threats is a short as possible. Because the decoys complete their missions in parallel, the objective can be formulated as minimising the longest positioning time of all the chosen decoy-threat pairs, where every threat is assigned to one decoy and all decoys are not assigned to more than one threat. The positioning time of decoy-threat pair $(i,j)$, denoted by $T^{\positioning}_{i,j}$, is defined by the earliest time at which decoy~$i$ can break the lock of threat~$j$ and remain indistinguishable from the asset thereafter. Furthermore, collisions between decoys must be avoided throughout. 

The task of positioning the decoys therefore involves a bottleneck assignment problem where the weights are determined by minimum-time trajectory planning problems with collision avoidance requirements. Simultaneously, optimising the task assignment and planning the motion of the individual decoys is challenging. The trajectory planning problems are coupled through the collision avoidance conditions that are non-convex with respect to decoy positions. Even without the coupling, evaluating the positioning times of all combinations of decoy-threat pairs is computationally intractable. We therefore first make the decision on the assignment of threats to decoys based on conservative estimates of positioning times that can be evaluated efficiently, $\hat{T}^\positioning_{i,j} \geq {T}^\positioning_{i,j}$, and then optimise the decoy trajectories with more complex motion planning given the assigned threats. To avoid collisions among assigned decoys, we apply local positions constraints that define a time-varying safe set, $\locCon_i(t,d) \subset \R^3$, for each decoy~$i$. By imposing the position constraints, $p_i(t) \in \locCon_i(t,d)$, for all decoys, $i \in \decoys$, we decouple the trajectory planning of the individual decoys. To this end, static target locations for jamming each threat have to be derived such that the decoy-threat allocation can be modelled as an assignment problem for which the tools introduced in~\cref{sec:pre_assignment,sec:pre_CollAvoid} are applicable.   

We then consider the motion control required to position an individual decoy $i\in\decoys$ such that it can start seducing its assigned threat $j\in\threats$ within a given time horizon $T$. We call control inputs \emph{admissible} if the resulting decoy states do not violate any state constraints. Inputs are called \emph{safe} if they lead to the decoy remaining in its local position constraint set, $p_i(t)\in\locCon_i(t,d)$, throughout the time horizon, $t\in[0,T]$. Finally, inputs are called \emph{satisfying} if they lead to a state trajectory for which at some instance within the time horizon, $T^\positioning_{i,j}\in[0,T]$, the decoy lies in the tracking cone, $p_i(T^\positioning_{i,j})\in\TrackingCone_j(T^\positioning_{i,j})$, while the threat is outside of the burn-through range, $\|y(T^\positioning_{i,j})-z_j(T^\positioning_{i,j})\|_\infty\geq\btrange_{i,j}(T^\positioning_{i,j})$, and the fake asset is indistinguishable from the true asset thereafter, $|\DopplerShift_{i,j}(t) - \DopplerShift_{\asset,j}(t)| \leq \maxdsvar$, for all $t\in[T^\positioning_{i,j},T]$. The objective in the motion planning is to find a strategy that generates control inputs that are admissible, safe, and satisfying and minimises the predicted positioning time, $T^\positioning_{i,j}$. 

\subsection{Motion Planning for Individual Decoy} \label{sec:MotionPlanning}
To plan the trajectory of decoy $i \in \decoys$, we use an approximate discrete-time model with a state at time step $k \in \{0,1,\dots,N\}$ defined by $x_i[k] = (x^p_i[k],x^v_i[k]) := ((p_i(kT_s),v_i(kT_s))$. The number of time steps considered is $N := \left\lceil\frac{T}{T_s}\right\rceil$, where $T_s$ is the sampling time. Considering only positions above ground, $    p_i(t)\in\StateConstraints_p:=\left\{p\in\R^3\,\big|\, [p]_3 \geq \frac{d}{2} \right\}$, and bounded velocities, $v_i(t)\in\StateConstraints_v:=\big\{v\in\R^3\,\big|\,\|v\|_\infty\leq\vmax \big\}$, for all $t\in[0,T]$, we assume the existence of a low-level controller that is capable of actuating the decoy such that it can be modelled as a first-order discrete time system with a one-step input delay. The decoy planning model, $x_i[k+1] = f_i(x_i[k],u_i[k],w_i[k])$, is then given by 
\begin{align}
    \label{eq:DiscreteModel}
     p_i((k+1)T_s) = p_i(kT_s) + T_s v_i(kT_s) + w^p_i[k]\,, \qquad
    v_i((k+1)T_s) = u_i[k] + w^v_i[k]\,,
\end{align}
with commanded velocity as input, $u_i[k]\in\InputConstraints := \big\{u\in\R^3\,\big|\|u\|_\infty\leq\vmax\,\big\}$, and uncertain disturbances, 
$w^p_i[k]\in\DisturbanceConstraints_p := \big\{w^p\in\R^3\,\big|\,\|w^p\|_\infty\leq\beta^p\big\}$, 
$w^v_i[k]\in\DisturbanceConstraints_v := \big\{w^v\in\R^3\,\big|\,\|w^v\|_\infty\leq\beta^v\big\}$, for $k \in \{0,1,\dots,N - 1\}$, where $\beta^p,\beta^v\geq 0$. We note that the state constraint set, $\StateConstraints := \StateConstraints_p \times \StateConstraints_v$, input constraint set, $\InputConstraints$, and disturbance support set, $\DisturbanceConstraints := \DisturbanceConstraints^p \times \DisturbanceConstraints^v$, can all be expressed as polyhedral sets. 

\begin{rem}
 Given the position and velocity of the decoy at a discrete time step, the commanded input and planning model in~\eqref{eq:DiscreteModel} can be interpreted as providing a set of target positions and velocities for the low-level controller to reach in the next time step. This is similar to the concept of waysets introduced in \cite{Shekhar2015}.
\end{rem}
 
For assigned threat $j \in \threats$, we formulate the planning of the decoy motion from time steps $k$ to $N$ with the following optimisation framework that we call the \ac{mptp}, 
  \begin{align}\label{eq:OptimalPositioningProblem}
  \operatorname{\ac{mptp}}_{i,j}[k]: &&
    \minimise{\uv_i^{N-1-k}[k]}&\qquad 
    \left.\begin{array}{l}T^\positioning_{i,j}\end{array}\right.\label{eq:OpPoPr_objective}\\
    &&\st  &\qquad \left.\begin{array}{l}\uv_i^{N-1-k}[k]\in\InputConstraints^{N-k}\,,\end{array}\right. 
    \xv^{N-k}_i[k] = F_i\left(x_i[k],\uv_i^{N-1-k}[k], \wv^{N-1-k}_i[k]\right)\,, \nonumber
    \\ &&&\qquad \left. 
    \begin{array}{l}
     \xv^{N-k}_i[k]\in\StateConstraints^{N+1-k}\,,\\
     \xv^{N-k}_i[k]\in\SafeConstraint_i[k]\,,\\
     \xv^{N-k}_i[k]\in\SpecConstraint_j[k]\,, 
    \end{array}\right\} \: \forall \wv^{N-1-k}_i[k] \in\DisturbanceConstraints^{N-k}\,. \nonumber 
\end{align}
where $F\left(x_i[k],\uv_i^{N-1-k}[k], \wv^{N-1-k}_i[k]\right)$ is the aggregation of the dynamics given in~\eqref{eq:DiscreteModel}. The local position constraints for collision avoidance are incorporated via $\SafeConstraint_i[k] := \big\{ \mathbf{x}^{N-k}[k] \,\big|\, \forall l\in\{k,k+1,\dots,N\}: x^p_i[l] \in \locCon_i(lT_s,d) \big\}$. We impose the additional state constraint, $\SpecConstraint_j[k]$, to encode the satisfaction of the positioning specification. While the state at the time step $k$, $x_i[k]$, is assumed to be given, the commanded inputs, $\uv_i^{N-1-k}[k]$, are optimisation variables.

\begin{rem}
    In \cite{Frick2017LCSS,Sessa2018HSCC} disturbance feedback policies are considered that robustly satisfy temporal logic constraints where there is uncertainty in the parameters of the specification. Only uncertainty arising from the approximation of the controlled decoy dynamics in~\eqref{eq:DiscreteModel} are considered in the \ac{mptp}. To account for changes in the specification parameters, e.g., mismatches in the estimated threat trajectory, the \ac{mptp} can be solved in every time step with a shrinking horizon. 
\end{rem}

To incorporate the constraint, $\SpecConstraint_j[k]$, into an optimisation framework, as in \cite{Frick2017LCSS}, appropriate atomic propositions have to be designed such that they capture the various subtasks of the positioning task. A suitable cost function that captures the number time step it takes to complete the positioning, $\EndStep_{i,j}:=\left\lceil\frac{T^\positioning_{i,j}}{T_s}\right\rceil$, is also required.

\section{Target Assignment and Local Safe Sets}\label{sec:TaskAssignment}
For high-level planning, we estimate the time it takes to position a decoy, $i\in\decoys$, for the seduction of a threat, $j\in\threats$, through the distance of the initial decoy position to a static target location, $g_j$, that is known to be suitable for jamming the threat. Given the time delay in the modelled decoy dynamics in~\eqref{eq:DiscreteModel} and assuming the initial decoy velocity is zero, the estimated positioning time is 
\begin{align}\label{eq:PositioningTimeEstimate}
   \hat{T}_{i,j} := \frac{\|g_j-p_i(0)\|_\infty}{\vref} + T_s \approx T^\positioning_{i,j}\,,
\end{align} where $\vref := \vmax - \beta^v$ is the maximum component-wise velocity that robustly satisfies the input constraints. 

\subsection{Nominal Jamming Locations}\label{sec:TargetPoints}
For successfully breaking the lock of the threat, the target jamming location has to lie in the tracking cone at the time the decoy reaches it, i.e., $g_j\in\TrackingCone_j(\hat{T}_{i,j})$. Furthermore, the threat has to be outside of the burn-through range at that time, i.e., 
 $
 \|y(\hat{T}_{i,j})-z_j(\hat{T}_{i,j})\|_2 \geq \btrange_{i,j}(\hat{T}_{i,j}) =\elconst_j\sqrt{\|g_j-z_j(\hat{T}_{i,j})\|_2} 
$.
Because the speed of a threat is much greater than the speed of the asset, we can approximate the path of an undisturbed threat by a straight line.
\begin{assum}\label{assum:StraightThreat}
The asset is stationary, i.e., $ \|v^\asset\| = 0$.
\end{assum}

The decoy may be far away from the tracking cone of the threat at time $t=0$. Therefore, a suitable target for the decoy is a static location that satisfies both the tracking cone and the no-burn-through conditions for as long as possible. 
\begin{thm}[Proof in Appendix~\ref{app:ProofTargetLocation}]\label{thm:TargetLocation}
Given the initial asset position, $y(0)$, a threat, $j \in \threats$, with initial position, $z_j(0)$, the jamming parameter, $\elconst_j$, and \cref{assum:StraightThreat}. The target location, 
\begin{align}\label{eq:targetpoint}
    g_j = y(0) - \frac{\elconst_j^2}{4{\|y(0)-z_j(0)\|_2}}\left(y(0)-z_j(0)\right)\,,
\end{align} is in the tracking cone, $g_j \in \TrackingCone_j(t)$, for all $t \in [0,\Tv_{j}] $ and if a decoy $i \in \decoys$ takes this position at time $t \in [0,\Tv_{j}] $, i.e, $p_i(t) = g_j$, then no burn through occurs at time $t$, $\|y(t) - z_j(t)\|_2 \geq \btrange_{i,j}(t)$, where $\Tv_{j}$ is the largest possible value.
\end{thm}

Target location~$g_j$, given in~\eqref{eq:targetpoint}, is viable for decoy~$i$ to jam
threat~$j$ if the decoy can reach it is before the threat does,
\begin{align}\label{eq:targetviability}
    \hat{T}_{i,j} < \Tv_{j} := \frac{4\|y(0)-z_j(0)\|_2-\elconst_j^2}{4\SpeedThreat}\,.
\end{align} 
It follows that if \eqref{eq:targetviability} is satisfied, moving with maximum component-wise velocity, $\vref$, towards the target jamming location, $g_j$, is a basic viable strategy for decoy $i$ to position itself for the seduction of threat $j$. 

\begin{rem}
    In~\cite{Shames2017CDC} an algorithm that numerically solves a nonlinear optimisation problem was proposed to find a target jamming location. \cref{thm:TargetLocation} provides analytic expression for a suitable target location if \eqref{eq:targetviability} holds. 
\end{rem}

\subsection{Threat Assignment}\label{sec:Assignment}
To determine which decoy defends against which threat, we use the estimated positioning times given in \eqref{eq:PositioningTimeEstimate}. In particular, the assignment of threats to decoys, denoted by $\Pi^*$, is required to minimise the largest estimated positioning time. This corresponds to a bottleneck minimising assignment introduced in \cref{defn:bottass}, i.e. 
\begin{align}\label{eq:bottReq}
\Pi^* \in \BAop_{\decoys,\threats}(\decoys\times\threats,\{\hat{T}_{i,j}\}_{(i,j)\in\decoys\times\threats})
    \,.
\end{align} 
We note that the estimated positioning times have an affine dependence on the distances between the decoys and the nominal jamming locations considering the infinity-norm distance function. To minimise the positioning times, we can therefore consider the assignment weights being given by the distances, $\dist = \{ \dist_{i,j} \}_{ (i,j) \in \decoys\times\threats }$, where $\dist_{i,j} := \vref (\hat{T}_{i,j} - T_s) = \| g_j - p_i(0) \|_\infty$. For this choice of weights we can apply the sequential bottleneck assignment method, outlined in~\cref{defn:lexass}, to obtain the collision avoidance properties described in \cref{sec:pre_CollAvoid}. The resulting assignment, $\Pi^* \in \LexAop_{ \decoys,\threats }( \dist )$, satisfies the the bottleneck requirement of \eqref{eq:bottReq}. The sequentially optimising assignment, $\Pi^*$, comes an $l$-th order bottleneck decoy-threat pair, $(i^*_l,j^*_l)$, and an $l$-th order robustness margin $\mu_l$, for all $l \in \{1,\dots,n\}$, that can be used to define local safe sets for all decoys if the following two assumptions hold. 

\begin{assum}\label{assum:ViableAssignment}
For all $l \in \{1,\dots,n\}$, the location, $g_{j^*_l}$, given by \eqref{eq:targetpoint}, is a viable position for decoy $i^*_l$ to jam threat $j^*_l$ according to the criterion given in~\eqref{eq:targetviability}, i.e., $\hat{T}_{i^*_l,j^*_l} < \Tv_{j^*_l}$. 
\end{assum} 

\begin{assum} \label{assum:asslexglobalsafedist} 
The robustness margins of all orders are greater than the diameter of the decoys, i.e., $\min_{ l \in \{1,\dots,n\} }\mu_l > d$.
\end{assum}

\cref{assum:asslexglobalsafedist} implicitly requires the spacing between initial decoy positions and nominal jamming locations to be sufficiently large. Given that this holds, the time-varying position set, $\locCon_i(t,d)$, introduced in~\cref{defn:LocalConstraint}, is non-empty and can be considered as a local safe set for any decoy, $i \in \decoys$. 
That is, according to~\cref{prop:CollisionAvoidance}, if all decoy satisfy their local position constraints at time $t$, i.e., $p_i(t) \in \locCon_i(t,d)$ for all $i \in \decoys$, then no decoys will collide with an assigned decoy at that time, i.e., $\|p_i(t)-p_i'(t)\|_\infty > d$ for all $i \in \{i^*_1,\dots,i^*_n\}$ and $i' \in \decoys \setminus \{i\}$. We note that the local safe sets do not guarantee the avoidance of collisions between two unassigned decoys. We assume that the coordination of unassigned decoys is less critical as long as they do not interfere with the assigned decoys and can therefore be addressed with separate strategies, e.g. by commanding unassigned decoys to remain at their initial positions.

\begin{rem}\label{rem:AssignmentComplexity}
Determining bottleneck edges and robustness margins of all orders has a complexity of $\bigO(mn^{3})$, see~\cite{Wood2020RAL,Khoo2020arXiv}. 
\end{rem}

\begin{rem}\label{rem:DistLex}
     In~\cite{Khoo2020arXiv} an algorithm finds a sequential bottleneck optimising assignment without centralised decision making. Agents only have knowledge of weights associated with them, e.g., decoy~$i\in\decoys$ knows the subset of weights $\weights_i:=\{\dist_{i,j}\,|\,j\in\threats\}$, and communicate local estimates of maximum and minimum global weights to other agents. 
\end{rem}

The local safe set for an assigned decoy, $i^*_l \in \{i^*_1,\dots,i^*_n\}$, consists of the intersection of two cubes that are centred at the initial decoys position, $p_{i^*_l}(0)$, and the assigned target destination, $g_{j^*_l}$, respectively. The lengths of the cube edges are given by $2(\ds_{i^*_l}(t,d) - \epsilon)$ and $2(\dt_{i^*_l}(t,d) - \epsilon)$, respectively, and are parametrised by the time-varying coordination parameter, $\ds(t),$ that is shared by all decoys as defined in~\eqref{eq:localBounds}. We let $\ds(t) = \ds^{\textrm{min}} := \frac{1}{2} \left( \mumin - d \right)$ for $t\leq Ts$ and increase with a constant rate, $ \ds(t) = \vref t + \ds^{\textrm{min}} $, for $t > T_s$. For $t \in [0,\hat{T}_{i^*_l,j^*_l }]$, the resulting safe set, $\locCon_{i^*_l}(t,d)$, includes the corresponding point on the straight-line maximum-speed trajectory from $p_{i^*_l}(0)$ to $g_{j^*_l}$. For a decoy that is not assigned to any threat, $i' \in \decoys \setminus \{i^*_1,\dots,i^*_n\}$, the local safe set is a single cube centred at $p_{i'}(0)$ with cube length $2(\ds_{i^*_n}(t,d) - \epsilon)$. An example and visualisation of the safe sets are presented in \cref{sec:CaseStudy}. We note that the local position constraints can be expressed as polyhedral sets. 
 
\section{Specification Satisfaction}\label{sec:specification}
The task of decoy~$i\in\decoys$ to position itself for the seduction of threat~$j\in\threats$ can be decomposed into subtasks each corresponding to set reachability problems. Combinations of such subtasks can in turn be described by \ac{ltl} formula as defined in \cref{sec:pre_LTL}. We define three atomic propositions, ${ap}^\cone_{j}$, ${ap}^\bt_{j}$, and ${ap}^\doppler_{j}$, by modelling the associated sets in \cref{sec:TrackingCone,,sec:BurnThrough,,sec:DopplerShift}, to capture the events of decoy~$i$ being in the tracking cone of threat $j$, threat~$j$ being in the burn-through range, and decoy~$i$ causing a tolerable Doppler shift, respectively. We combine the atomic propositions to an \ac{ltl} formula in~\cref{sec:PositioningTask} and discuss how to encode it in \cref{sec:MIencoding}. 

\subsection{Decoy in Tracking Cone}\label{sec:TrackingCone}
The set of positions in the tracking cone of threat $j$ at time $t$ is defined by non-linear constraints in~\eqref{eq:trackingcone}. We make an inner approximation of the tracking cone, $\ApproxTrackingCone_j(t) \subset \TrackingCone_j(t)$, with affine inequalities. In this way, we can define the atomic proposition, ${ap}^\cone_{j}$, that if satisfied, i.e., $\xv_i^{N-k}[k] \satisfies {ap}^\cone_{j}$, implies that decoy $i$ is located in the tracking cone of threat $j$ at time step $k$, i.e., $p_i(kT_s) \in \TrackingCone_j(kT_s)$.   

\begin{figure}%[th]
    \centering
    \begin{subfigure}{0.24\linewidth}
      \centering
      \includegraphics[width=\linewidth]{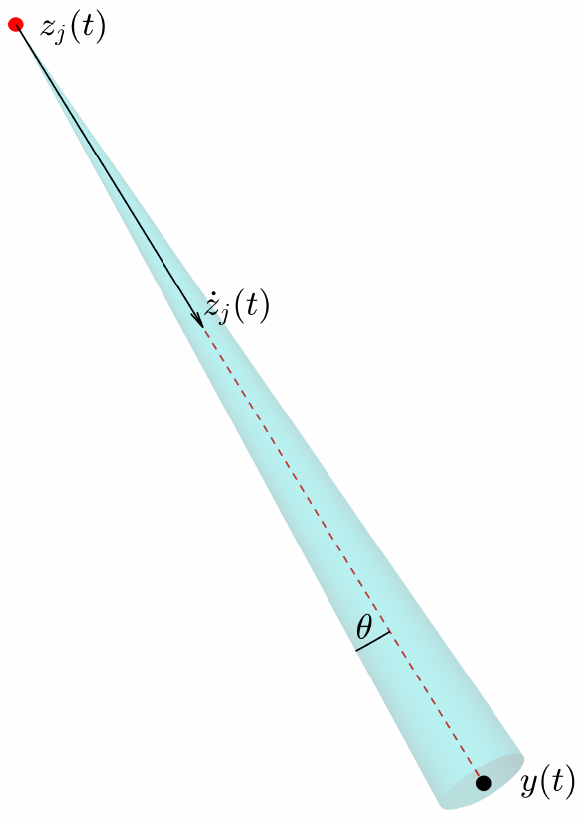}
      \caption{$\TrackingCone_j(t)$}
        \label{fig:TrackingCone_pure}
    \end{subfigure}
        \begin{subfigure}{0.24\linewidth}
      \centering
      \includegraphics[width=\linewidth]{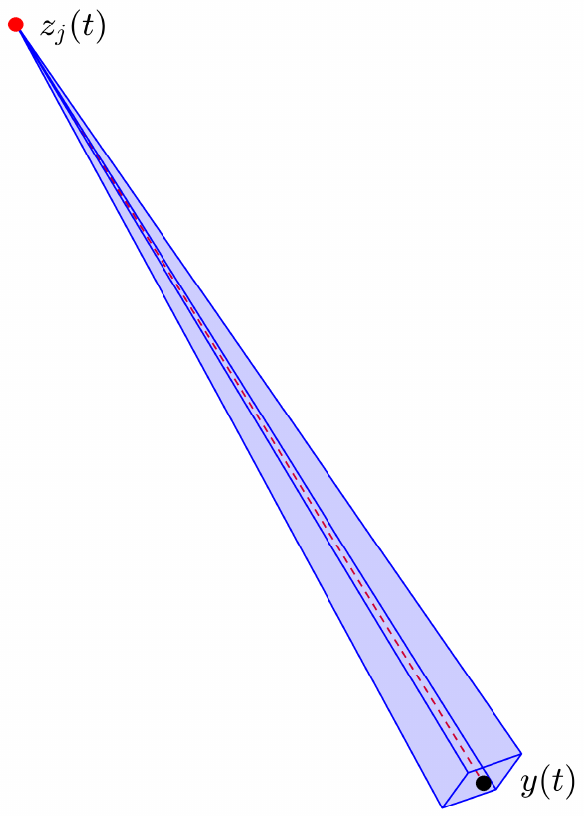}
      \caption{$\ApproxTrackingCone_j(t)$} 
        \label{fig:TrackingCone_approx}
    \end{subfigure}
    \begin{subfigure}{0.24\linewidth}
      \centering
      \includegraphics[width=\linewidth]{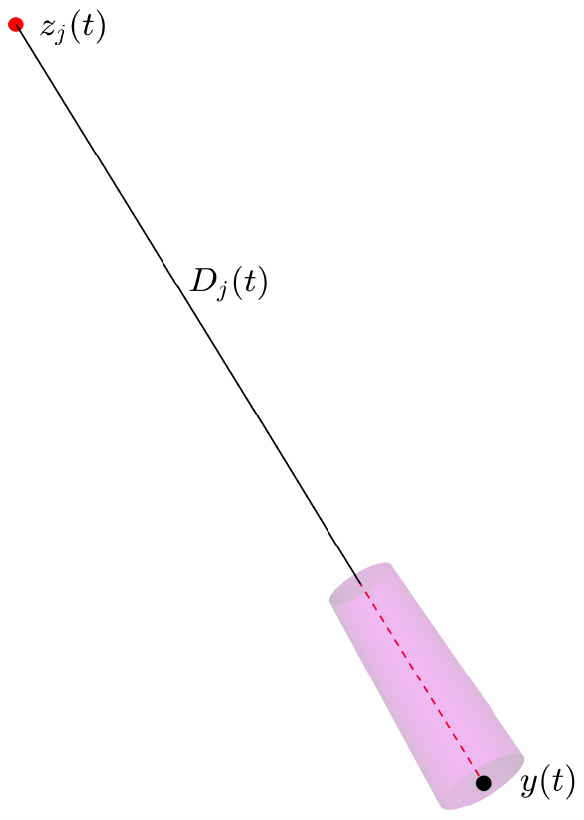}
      \caption{$\TrackingCone_j(t) \cap \BurnThroughSet_j(t)$} 
        \label{fig:TrackingCone_burnthrough}
    \end{subfigure}
        \begin{subfigure}{0.24\linewidth}
      \centering
      \includegraphics[width=\linewidth]{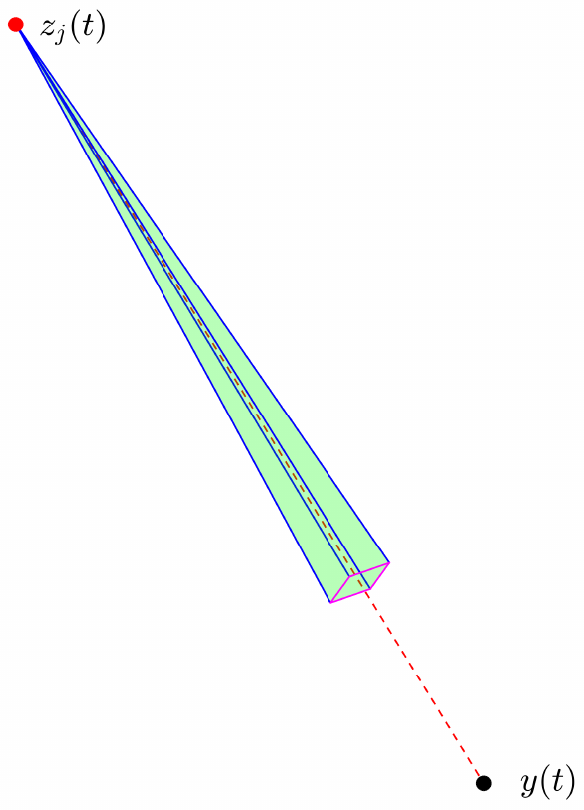}
      \caption{$\ApproxTrackingCone_j(t) \setminus \ApproxBurnThroughSet_j(t)$} 
        \label{fig:TrackingCone_hitset}
    \end{subfigure}
      \caption{Illustration of the tracking cone, $\TrackingCone_j(t)$, the region for which burn through occurs if a decoy is positioned in it, $\BurnThroughSet_j(t)$, and their approximations for threat~$j$ %, at position $z_j(t)$  (red dot), 
      heading towards the asset.}%, $\asset$, at position $y(t)$ (black dot).}
    \label{fig:Propositions}
\end{figure}

We approximate the tracking cone, with aperture $2\theta$, by the union of four half-spaces. The normal vectors of the boundary planes, $b^{1+}_j(t), b^{1-}_j(t), b^{2+}_j(t), b^{2-}_j(t)\in\R^3$, are obtained by rotating the cone axis, $\dot{z}_j(t)$, around two rotation axis, $\rot^1,\rot^2\in\R^3$, that are orthogonal to each other, ${\rot^1}^\top \rot^2=0$, and the cone axis ${\rot^1}^\top \dot{z}_j(t) = {\rot^2}^\top \dot{z}_j(t) = 0$, by angles $\RotAng^+ = \left(\aperture + \frac{\pi}{2}\right)$ and $\RotAng^- =-\left(\aperture + \frac{\pi}{2}\right)$. We approximate the distance constraint in~\eqref{eq:trackingcone} with an additional half-space constraint with boundary perpendicular to $\dot{z}_j(t)$ at a distance of $\|y(t)-z_j(t)\|_2\cos(\aperture)$ from $z_j(t)$. With $B_j:=\begin{bmatrix} b^{1+}_j(t) & b^{1-}_j(t) & b^{2+}_j(t) & b^{2-}_j(t) \end{bmatrix}^\top$, the resulting polyhedral approximation of the tracking cone is, 
\begin{align}\label{eq:approxtrackingcone}
  \ApproxTrackingCone_j(t) := \big\{p\in\R^3\,\big|\, B_j p &\leq B_j z_j(t),\, \dot{z}_j(t)^\top p \leq \dot{z}_j(t) ^\top z_j(t)  + \|y(t)-z_j(t)\|_2\SpeedThreat\cos(\aperture)
  \big\} \,.
\end{align}
\cref{fig:TrackingCone_approx} shows an example of $\ApproxTrackingCone_j(t)$.
The atomic proposition, ${ap}^\cone_{j}$, is defined by the time-step dependent polyhedral set, $\polytope^\cone_{j}[k] %= \big\{ x\in\R^{n_x} \,\big|\, \Phi_{{ap}_j^\cone}[k]x \leq \Psi_{{ap}_j^\cone}[k] \big\}
% :=\bigg\{
% \begin{bmatrix}
%   p^\top & v^\top
% \end{bmatrix}^\top\in\R^{n_x} \,\bigg|\, p\in\ApproxTrackingCone_j(kT_s) 
:=\big\{(p,v) \,\big|\, p\in\ApproxTrackingCone_j(kT_s) 
\big\} $.

\subsection{Threat in Burn-Through Range}\label{sec:BurnThrough}
Threat~$j$ has to be outside of the burn-through range, defined in \eqref{eq:BurnThroughRange}, for decoy~$i$ to be able to break its lock. Conversely, if at time $t$ the decoy is located so far away from the threat such that $ \|p_i(t) - z_j(t)\|_2 > \DecoyRange_j(t) := \frac{1}{\elconst_j^2} \|y(t) - z_j(t)\|_2^2 $, then the threat is in the burn-through range at time $t$. The set of decoy positions for which the threat is within the burn-through range is $\BurnThroughSet_j(t) := \big\{p\in\R^3\,\big|\,\|p-z_j(t)\|_2>D_j(t)\}$. \cref{fig:TrackingCone_burnthrough} show and illustration of $\TrackingCone_j(t) \cap \BurnThroughSet_j(t)$. To define the atomic proposition, ${ap}^\bt_j$, we introduce a set, $\ApproxBurnThroughSet_j(t)$, with a single affine constraint, such that the set of decoy positions in the tracking cone for which the threat is in the burn-through range is approximated from outside, i.e., $\TrackingCone_j(t) \cap \ApproxBurnThroughSet_j(t) \supset \TrackingCone_j(t) \cap \BurnThroughSet_j(t)$.
Assuming the decoy is in the tracking cone at time step $k$, i.e., $\xv_i^{N - k}[k]\satisfies {ap}_j^\cone$, it is guaranteed that the threat is not in the burn-through range at time step $k$, i.e. $\|y(kT_s)-z_j(kT_s)\|_2 \geq \burnthrough_{i,j}(kT_s)$, if ${ap}^\bt_j$ is not satisfied, i.e., $\xv_i^{N - k}[k] \satisfies \lnot {ap}^\bt_j$. \cref{fig:TrackingCone_hitset} shows an example of the corresponding target set $\ApproxTrackingCone_j(t) \setminus \ApproxBurnThroughSet_j(t)$.

To make the outer approximation, we consider the projection of the difference vector between the decoy and the threat positions onto the cone axis, $p'_i(t) := \frac{1}{s^z}\left(p_i(t)-z_j(t)\right)^\top\dot{z}_j(t)$. Given that the decoy is in the tracking cone of the threat at time $t$, i.e., $p_i(t)\in\TrackingCone_j(t)$, we necessarily have $p'_i(t) > D_j(t)\cos(\aperture)$ if the threat is in the burn-through range. With an arbitrarily small, $\epsilon>0$, the set of positions that satisfy this condition forms a half space and is given by
\begin{align}\label{eq:approxburnthrough}
  \ApproxBurnThroughSet_{j}(t):=\big\{p\in\R^3\,\big|\,-\elconst_j^2\dot{z}_j(t)^\top p \leq -\elconst_j^2\dot{z}_j(t) ^\top z_j(t)  - \|y(t)-z_j(t)\|_2^2\SpeedThreat\cos(\aperture) - \epsilon\big\}\,.
\end{align}
The atomic proposition, ${ap}^\bt_j$, is defined by the time-step dependent polyhedral set, 
$\polytope^\bt_j[k] 
%=\big\{ x \in \R^{n_x} \,\big|\, \Phi_{{ap}_j^\bt}[k] x \leq \Psi_{{ap}_j^\bt}[k] \big\} 
% := \bigg\{
% \begin{bmatrix}
%   p^\top & v^\top
% \end{bmatrix}^\top \in \R^{n_x} \,\bigg|\, p \in \ApproxBurnThroughSet_j(t)\bigg\}
:= \big\{(p,v) \,\big|\, p \in \ApproxBurnThroughSet_j(kT_s)\big\}
$.
% %\end{align} 

\subsection{Bounded Doppler Shift}\label{sec:DopplerShift}
For threat~$j$ to be unable to distinguish between the fake asset, $\FakeAsset_i$,  portrayed by decoy~$i$ and the true asset, $\asset$, the Doppler shift caused by the return signal of the decoy has to be limited. Specifically, the decoy velocity in direction of the threat velocity is required to be bounded. With~\cref{assum:StraightThreat} and~\eqref{eq:ds}, the Doppler shift condition is equivalent to the decoy velocity being in the time-varying polyhedral set,
\begin{align}\label{eq:dopplerconstrained}
    \lowshift_j(t):=\bigg\{v\in\R^3 \,\bigg|\, -\dot{z}_j(t)^\top v \leq -\dot{z}_j(0)^\top v^\asset + \frac{\maxdsvar\SpeedThreat c}{\TransFreq},\, \dot{z}_j(t)^\top v \leq \dot{z}_j(0)^\top v^\asset + \frac{\maxdsvar\SpeedThreat c}{\TransFreq}\bigg\}\,.\nonumber
\end{align} We define ${ap}^\doppler_j$ to be the atomic proposition of satisfying the Doppler-shift condition of threat~$j$. If this proposition is satisfied by the decoy, $\xv_i^{N-k}[k] \satisfies {ap}^\doppler_j$, then, from the perspective of the threat, the Doppler shift caused by the decoy is indistinguishable from the one caused by the asset at time step $k$, i.e., $|\DopplerShift_{i,j}(kT_s) - \DopplerShift_{\asset,j}(kT_s)| \leq \maxdsvar$. The polyhedral set associated with ${ap}^\doppler_j$ is 
$
\polytope^\doppler_j[k] 
%= \big\{ x \in\R^{n_x} \,\big|\, \Phi_{{ap}_j^\doppler}[k] x \leq \Psi_{{ap}_j^\doppler[k]} \big\} 
% := \bigg\{
% \begin{bmatrix}
% p^\top & v^\top  
% \end{bmatrix}^\top
% \in \R^6 \,\bigg|\, v \in \lowshift_j(kT_s)\bigg\}
:= \big\{(p,v) \,\big|\, v \in \lowshift_j(kT_s)\big\}
$.

\subsection{Specification of Decoy Positioning Task}\label{sec:PositioningTask}
The task of positioning decoy $i$ for the seduction of threat $j$ requires the combination of the subtasks of reaching the tracking cone, i.e. satisfying ${ap}^\cone_j$, avoiding burn through, i.e. satisfying $\lnot{ap}^\bt_j$, and bounding the Doppler shift, i.e. satisfying ${ap}^\doppler_j$.  We apply the \ac{ltl} semantics introduced in Definition~\ref{defn:ltl} and the atomic propositions ${ap}^\cone_j$, ${ap}^\bt_j$, and ${ap}^\doppler_j$ to define the \emph{positioning specification}, $\varphi_j^\positioning$. The propositions ${ap}^\cone_j$ and $\lnot{ap}^\bt_j$ have to be satisfied at the same time in order to break the lock of the threat. From the moment the lock is broken, $\lalways{ap}^\doppler_j$ has to be satisfied. The positioning should be completed before the end of the planning horizon but the specific completion time within the horizon should not be fixed. We, therefore, define
\begin{align}\label{eq:PositioningSpecification}
    \varphi_j^\positioning:=\leventually \left({ap}^{\cone}_j \land \lnot {ap}^\bt_j \land \lalways {ap}^\doppler_j\right)\,.
\end{align}  

We note that, if $x_i^{N-k}[k] \satisfies \varphi_j^\positioning$, there exists a time step within the horizon, $l \in \{k,k + 1,\dots,N\}$, at which the decoy can start the seduction process and switch to a different motion control strategy. By imposing a bounded Doppler shift and safety constraints for the remainder of the planning horizon, we ensure that there exist inputs that satisfy the crucial Doppler shift and collision avoidance conditions until at least the end of the planning horizon. For the discrete-time planning framework introduced in~\cref{sec:MotionPlanning}, the positioning task at time step $k$ can be formulated with the state constraint set 
$
      \SpecConstraint_j[k] 
      := \big\{\mathbf{x}^{N-k}[k] \,\big|\, \xv_i^{N-k}[k] \satisfies \varphi^\positioning_{j} \big\}
$.

\subsection{Mixed-Integer Encoding of Specification}\label{sec:MIencoding}
Because the atomic propositions are all defined via polyhedral sets, the satisfaction of the positioning specification, $\xv^{N-k}_i[k] \satisfies \varphi^\positioning_{j}$, can be written as a set of affine constraints, see \cite{Wolff2014ICRA}. To formulate these constraints, we introduce auxiliary variables of which some are restricted to binary values. We couple the auxiliary variables to the state of decoy~$i$ such that fixing the auxiliary variables to particular values enforces the positioning specification, $\varphi_j^\positioning$, to be satisfied. In contrast to the approach in \cite{Wolff2014ICRA}, we do not require the auxiliary variables to indicate if the specification is not satisfied. To this end, because the satisfaction of ${ap}_j^\cone$ and ${ap}_j^\doppler$ is of importance for the satisfaction of $\varphi_j^\positioning$, there should exist values of the associated auxiliary variables that enforce the decoy state being in the corresponding sets $\polytope_j^\cone[k]$ and $\polytope_j^\doppler[k]$. However, the state not being in these sets does not need to be reflected in the values of the auxiliary variables. On the other hand, because we need to enforce that ${ap}_j^\bt$ is not satisfied at some time steps in order to satisfy $\varphi_j^\positioning$, there should be values of associated auxiliary variables that imply that the state is not in $\polytope_j^\bt[k]$. However, the decoy state not being in  $\polytope_j^\bt[k]$ does not need to affect the auxiliary variables.

In~Appendix~\ref{sec:MIconstraints} we derive the specific constraints that encode the requirement of satisfying the positioning specification. This specific encoding differs from the standard bijective encoding of auxiliary variables to formula satisfaction introduced in \cite{Wolff2014ICRA}, requires fewer constraints, and enables the derivation of the minimum-time cost function discussed in \cref{sec:CostFunction}. The auxiliary variables include one binary variable per discrete time step for each half-space constraint used to defined the polyhedral sets associated with the atomic propositions. For motion planning at time step $k$, we can enforce the requirement of decoy~$i$ positioning itself for the seduction of threat~$j$, i.e, $\xv_i^{N-k}[k] \in \SpecConstraint_j[k]$, by condition $\gamma_{\varphi^\positioning_{j}}[k] = 1$, where $\gamma_{\varphi^\positioning_{j}}[k]$ is one of the introduced auxiliary variables. All the constraints that encode the specification are affine in the state, $\mathbf{x}_i^{N-k}[k]$, and the mixed-integer auxiliary variables, denoted collectively by $\omega \in \Omega :=\R^{n_c} \times \{0,1\}^{n_b}$, with $n_b = 8(N-k) + 8$ binary and $n_c=5(N-k) + 3$ continuous auxiliary variables.       

\section{Robust Minimum-Time Decoy Positioning}\label{sec:RobOpt}
We consider the \ac{mptp} at a discrete time step $k \in \{0,1,\dots,N-1\}$ for decoy~$i\in\decoys$ and threat~$j\in\threats$. Robustly admissible inputs guarantee the decoy satisfying the state constraints at every time step within the remaining planning horizon, $\xv_i^{N-k}[k]\in\StateConstraints^{N+1-k}$, for all realisations of the disturbances $\wv_i^{N-1-k}[k]\in\DisturbanceConstraints^{N-k}$. The set of robustly admissible inputs can be expressed as  
\begin{align}
    \RobAdmConstraint_i[k] :=
    %& 
    \big\{\uv \in \InputConstraints^{N-k} \, \big|\,\forall \wv \in \DisturbanceConstraints^{N-k} : \xi(x_i[k],\uv,\wv)\leq 0\big\}
\,,
    %\\ =&\big\{\uv\in\InputConstraints^{N-k} \,\big|\, \max_{\wv \in \DisturbanceConstraints^{N-k}} \: \max_{q\in\{1,\dots,n_X\}} [\xi(x_i[k],\uv,\wv)]_q\leq 0\big\}\,,
\end{align} 
where $\xi(x_i[k],\uv,\wv)$ results from the dynamics in~\eqref{eq:DiscreteModel}, 
has $n_X = 7(N-k) + 7$ rows, and is affine in $\uv$ and $\wv$. Robustly safe inputs guarantee that for all disturbances, $\wv_i^{N-1-k}[k]\in\DisturbanceConstraints^{N-k}$, the decoy is in the safe set, i.e., $p_i(lT_s)\in\locCon_i(lT_s,d)$, for every time step $l\in\{k,k+1,\dots,N\}$. The set of robustly safe inputs can be described by 
\begin{align}
    \RobSafeConstraint_i[k] :=%& 
    \big\{\uv\in\InputConstraints^{N-k} \,\big|\, \forall \wv \in \DisturbanceConstraints^{N-k} : \lambda_i(x_i[k],\uv,\wv)\leq 0\big\}
    %, %\\
     %=&\big\{\Uv\in\InputConstraints^N \,\big|\,\max_{\Wv\in\DisturbanceConstraints^N} \: \max_{q\in\{1,\dots,n_L\}} [\lambda_i(x_i(0),\Uv,\Wv)]_q\leq 0\big\}\,.
\,,
\end{align} 
with $\lambda_i(x_i[k],\uv,\wv)$ consisting of $n_L = 12(N - k) + 12$ rows that are affine in $\uv$ and $\wv$.

Robustly satisfying inputs lead to the decoy satisfying the positioning specification, $\xv^{N-k}_i[k] \in \SpecConstraint_j[k]$, for any disturbance, $\wv_i^{N-1-k}[k] \in \DisturbanceConstraints^{N-k}$. By applying the specification encoding of~Appendix~\ref{sec:MIconstraints}, the set of robustly satisfying inputs can be expressed as  
\begin{align}
    \RobSpecConstraint_{i,j}[k]:=%&
\big\{\uv\in \InputConstraints^{N-k} \,\big|\,  \forall \wv \in \DisturbanceConstraints^{N-k} : \exists \aux \in \Omega \textrm{ s.t. } 
    \sigma_j(x_i[k],\uv,\wv,\aux)\leq 0\big\}
    \,,
\end{align}
    %, 
    % \\    =&\big\{\Uv\in \InputConstraints^N \,\big|\, \max_{\Wv\in\DisturbanceConstraints^N} \: \min_{\aux\in\Omega} \: \max_{q\in\{1,\dots, n_S\}} \: [\sigma_j(x_i(0),\Uv,\Wv,\aux)]_q\leq 0 \big\}\,,
%\end{align}
where $\sigma_j(x_i[k],\uv,\wv,\aux)$ has $n_S=37(N-k)+33$ rows and is affine in $\uv$, $\wv$, and $\aux$.
Constraining $\uv^{N-1-k}_i[k]\in\RobSpecConstraint_{i,j}[k]$ involves the maximisation of $\sigma'_i(x_i[k],\wv) := \min_{\aux\in\Omega} \max_{q \in \{1,\dots,n_S\}} \: [\sigma_j(x_i[k],\uv,\wv,\aux)]_q$ with respect to $\wv$  which is difficult because $\sigma'_i(x_i[k],\wv)$ is not concave in $\wv$. As proposed in~\cite{Frick2017LCSS}, we switch the order of the quantifiers $\forall \wv\in\DisturbanceConstraints^N$ and $\exists \aux\in\Omega$ to obtain an inner approximation of $\RobSpecConstraint_{i,j}[k]$, i.e.,  
\begin{align}
    \ApproxRobSpecConstraint_{i,j}[k] := 
    %& 
    \big\{\uv\in \InputConstraints^N \,\big|\, \exists \aux\in\Omega \textrm{ s.t. } \forall \wv\in\DisturbanceConstraints^N : \sigma_j(x_i(0),\uv,\wv,\aux)\leq 0 \big\} %\\
    %
    %\big\{\uv\in \InputConstraints^N \,\big|\, \min_{\aux\in\Omega} \: \max_{\wv\in\DisturbanceConstraints^N} \:  \max_{q\in\{1,\dots, n_S\}} \: [\sigma_j(x_i(0),\uv,\wv,\aux)]_q\leq 0 \big\}
    \,.
\end{align}
The constraint $\uv^{N-1-k}_i[k] \in \ApproxRobSpecConstraint_{i,j}[k]$ is less difficult and sufficiently implies that the inputs are robustly satisfying because $\ApproxRobSpecConstraint_{i,j}[k]\subseteq\RobSpecConstraint_{i,j}[k]$. Given the constraint sets, $\RobAdmConstraint_i[k]$, $\RobSafeConstraint_j[k]$, and $\ApproxRobSpecConstraint_{i,j}[k]$, we derive a cost function in~\cref{sec:CostFunction} and in~\cref{sec:TractableProblem} formulate a computationally tractable approach for the \ac{mptp}. 

\subsection{Cost Function for Minimising Completion Time}\label{sec:CostFunction}
Decoy $i$ fulfils the task of positioning itself for the seduction of threat $j$, if the sequence of discrete decoy states starting at time step $k$ satisfies the specification $\varphi^\positioning_{j} = \leventually \bar{\varphi}_j$, with $\bar{\varphi}_j := {ap}^{\cone}_j \land \lnot {ap}^\bt_j \land \lalways {ap}^\doppler_j$. In this case, there exists a time step, $\bar{k}_{i,j} \in \{k,k+1,\dots,N\}$, for which the state sequence starting at time step $\bar{k}_{i,j}$ satisfies $\bar{\varphi}_j$, i.e., $\xv_i^{N-k}[k] \satisfies \varphi^\positioning_{j}$ implies that $\exists \bar{k}_{i,j} \in \{k,k+1,\dots,N\}\:\textrm{ s.t. }\:\xv_i^{N-\bar{k}_{i,j}}[\bar{k}_{i,j}] \satisfies \bar{\varphi}_j$. From the definition of the eventually operator in~\cref{sec:pre_LTL}, we know that if $\bar{\varphi}_j$ is satisfied at time step $\bar{k}_{i,j}>k$, then $\varphi^\positioning_{j}$ is also satisfied for previous time steps $l\in\{k,k+1,\dots,\bar{k}_{i,j}\}$, i.e., $\xv_i^{N-\bar{k}_{i,j}}[\bar{k}_{i,j}] \satisfies \bar{\varphi}_j$ implies that $\forall l\in\{k,k+1,\dots,\bar{k}_{i,j}\} \: : \: \xv_i^{N-l}[l] \satisfies \varphi^\positioning_{j}$. The encoding in~\cref{sec:MIencoding} guarantees that $\xv_i^{N-k}[k] \satisfies \varphi^\positioning_{j}$ if $\gamma_{\varphi^\positioning_{j}}[k]=1$ but the other direction is not enforced, i.e., $\gamma_{\varphi^\positioning_{j}}[k]=0$ does not imply that $\xv_i^{N-k}[k] \not\satisfies \varphi^\positioning_{j}$.  Given that $\gamma_{\varphi^\positioning_{j}}[k]=1$, the sum of the time steps $l \in \{k,k+1,\dots,N\}$ where $\gamma_{\varphi^\positioning_{j}}[l]=1$ is an upper bound on the number of time steps required after $k$ to complete the specification $\varphi^\positioning_{j}$. Therefore, the time step at which the specification is first completed, $N^\positioning_{i,j}$, can be bounded by  
\begin{align}\label{eq:ApproxPositioningTime}
    \hat{N}^\positioning_{i,j}[k] := k - 1 + \sum_{l=k}^N \gamma_{\varphi^\positioning_{j}}[l] \geq N^\positioning_{i,j}\,,
\end{align} if $\gamma_{\varphi^\positioning_{j}}[k]=1$. We note that $\hat{N}^\positioning_{i,j}[k]$ is an affine function of the auxiliary mixed-integer variables, $\aux\in\Omega$. 

\subsection{Tractable Motion Planning}\label{sec:TractableProblem}
With the inner-approximation of the robustly satisfying input set, $\ApproxRobSpecConstraint_{i,j}[k]$, and the upper bound on the positioning time step, $\hat{N}^\positioning_{i,j}[k]$, we formulate the following approximation of $\operatorname{\ac{mptp}}_{i,j}[k]$ defined in~\cref{sec:MotionPlanning},
\begin{align}\label{eq:ApproxOptimalPositioningProblem}
    \widehat{\operatorname{\ac{mptp}}}_{i,j}[k]: &&\minimise{\uv, \aux} & \qquad \hat{N}^\positioning_{i,j}[k] T_s
    \\
    &&\st & \qquad \uv\in\InputConstraints^{N-k},\: \aux\in\Omega \,, \nonumber%\label{eq:ApproxOptimalPositioningProblem_feasible1}
    \\
    &&& \qquad \max_{\wv\in\DisturbanceConstraints^{N-k}} \: [\xi(x_i[k],\uv,\wv)]_q \leq 0 &\forall q \in \{1,\dots,n_X\}\,,
    \\
    &&& \qquad \max_{\wv\in\DisturbanceConstraints^{N-k}} \: [\lambda_i(x_i[k],\uv,\wv)]_q \leq 0 & \forall q \in \{1,\dots,n_L\} \,, 
    \\
    &&& \qquad\max_{\wv\in\DisturbanceConstraints^{N-k}} \: [\sigma_j(x_i[k],\uv,\wv,\aux)]_q \leq 0  & \forall q \in \{1,\dots,n_S\}\,,
    %\} \: %\forall \wv\in\DisturbanceConstraints^{N-k}\,. \nonumber%\label{eq:ApproxOptimalPositioningProblem_feasible2}
\end{align} 
where the mixed-integer auxiliary variables that couple the decoy state to the positioning specification are included as optimisation variables. We note that if inputs, $\uv_i^{N-1-k}[k]$, are in the feasible set of the approximated problem, $\widehat{\operatorname{\ac{mptp}}}_{i,j}[k]$, then they are also in the feasible set of the original problem, $\operatorname{\ac{mptp}}_{i,j}[k]$. Therefore, the optimal value of $\widehat{\operatorname{\ac{mptp}}}_{i,j}[k]$ is an upper bound on the optimal value of $\operatorname{\ac{mptp}}_{i,j}[k]$, i.e., $\hat{N}^{\positioning*}_{i,j}[k] T_s \geq T^\positioning_{i,j}$. Moreover, $\widehat{\operatorname{\ac{mptp}}}_{i,j}[k]$ can be formulated as a \ac{milp} by invoking linear programming duality and solved using off-the-shelf solvers as shown in \cite{Bertsimas2011SIAMr,Frick2017LCSS}. It therefore provides a tractable approach to the decoy motion planning problem. However, the success of the motion planning relies on the existence of low-level control that can actuate the decoy to follow the optimised commands obtained based on the simplified discrete-time planning model. 

\begin{assum}\label{assum:IntersampleBahaviour}
Given values of the optimisation variables, $(\hat{\uv},\hat{\aux})$, that fulfil the constraints of $\widehat{\operatorname{\ac{mptp}}}_{i,j}[k]$. For all discrete time steps $l \in \{k,\dots,N-1\}$, there exists low-level control action such that $(p_i(t),v_i(t)) \in \StateConstraints$ and $p_i(t) \in \locCon_i(t,d)$ for all $t \in \big[lT_s,(l+1)T_s\big)$. Furthermore, for any $l' \in \{k,\dots,N-1\}$ for which ${\gamma}_{\lalways {ap}^\doppler_j}[l']=1$, where ${\gamma}_{\lalways {ap}^\doppler_j}[l']$ is the element of $\hat{\aux}$ corresponding to the encoding of the formula $\lalways{ap}^\doppler_j$ at time step $l'$ defined in~\eqref{eq:gammaalwaysdoppler}, there exists low-level control action such that $v_i(t) \in \lowshift_j(t)$ for all $t\in \big[l'T_s,(l'+1)T_s\big)$.
\end{assum}

\section{Case Study}\label{sec:CaseStudy}
\begin{figure}%[th]
    \centering
    \begin{subfigure}{0.49\linewidth}
      \centering
      \includegraphics[width=\linewidth]{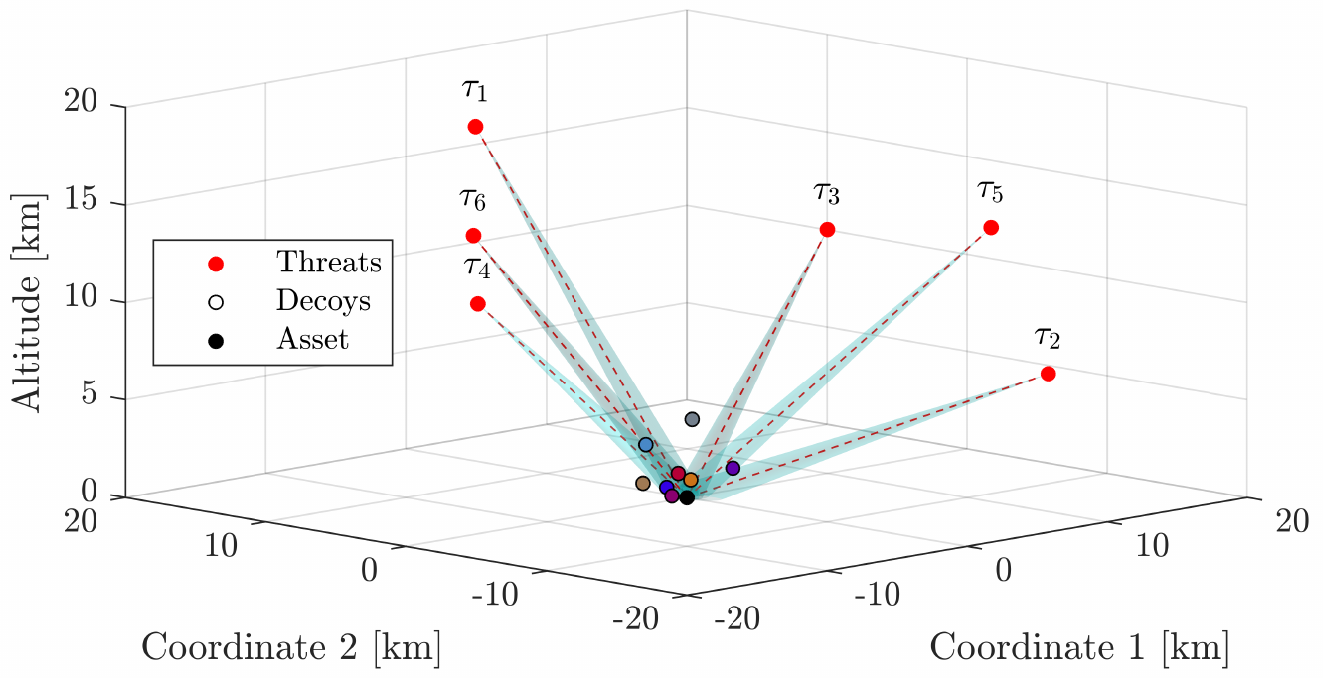}
      \caption{Overview of threats} 
        \label{fig:InitialOverview}
    \end{subfigure}
        \begin{subfigure}{0.49\linewidth}
      \centering
      \includegraphics[width=\linewidth]{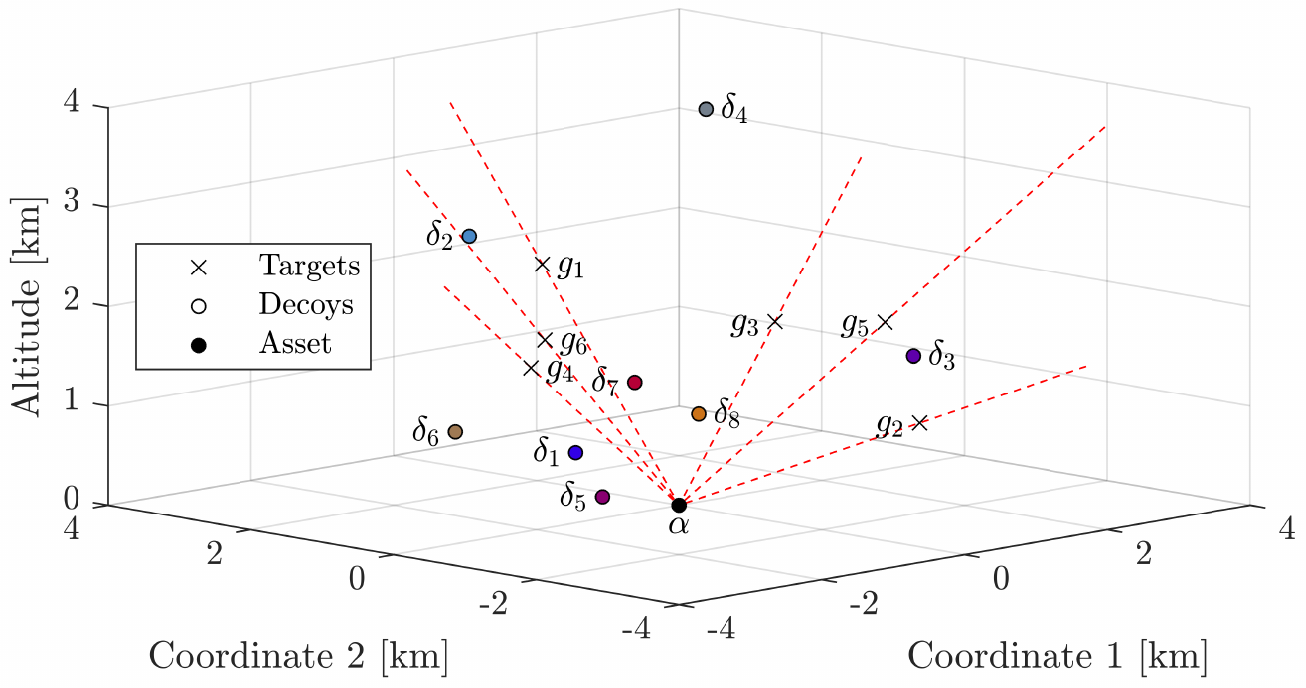}
      \caption{Close-up view of decoys and target jamming locations} 
        \label{fig:InitialCloseUp}
    \end{subfigure}
      \caption{Initial configuration of the threats, decoys, and asset at time $t=0$s.}
    \label{fig:Initial}
\end{figure}

We apply the positioning strategy to a scenario with $m=8$ decoys, $\decoys=\{\delta_1,\dots,\delta_{8}\}$, and $n=6$ threats, $\threats=\{\tau_1,\dots,\tau_6\}$. The initial configuration of decoy positions, $p_i(0)$, $i \in \decoys$, threat positions, $z_j(0)$, $j \in \threats$, and the asset position, $y(0)$, is shown in \cref{fig:Initial}.  The threats are travelling towards the asset with a speed of $\SpeedThreat=274$m/s and are positioned at distances between 19km and 23km from the asset at time $t=0$. The decoys are initialised with zero velocity and positions within in a distance of 4km from the asset.  The decoy diameters are assumed to be smaller than $d = 2$m. The maximum component-wise velocity the decoys can fly is $\vmax = 40$m/s.  We assume that the low-level controller can satisfy the discrete-time planning model constraints given in \eqref{eq:DiscreteModel} with a sampling time of $T_s = 2$s and uncertainty bound values of $\beta^p = 6$m and $\beta^v = 1$m/s. 
The tracking cone aperture is assumed to be $2 \aperture = 4 \degree$, the radar frequency is $F = 1$GHz, the jamming constant is $\elconst_j = \elconst=105\sqrt{\textrm{m}}$, for all threats $j\in\threats$, and the maximum tolerated Doppler shift deviation is $\maxdsvar=50$Hz. 

\subsection{Task Assignment and Collision Avoiding Safe Sets}
The points along the estimated flight path of each threat at which breaking the lock is achievable for the longest possible time are determined as described in~\cref{sec:TargetPoints}. The resulting target jamming locations, $g_j$, for all threats, $j\in\threats$, are shown as black crosses in~\cref{fig:InitialCloseUp}. The infinity-norm distances between the initial decoy positions and the target jamming locations, $\dist_{i,j}=\|g_j - p_i(0)\|_\infty$, $(i,j)\in \decoys\times\threats$, are used as weights to determine which one of the decoys is assigned to each threat. The resulting sequentially bottleneck optimising assignment, $\Pi^* \in \LexAop_{\decoys,\threats}(\dist)$, is presented in \cref{tab:assignment}. The reference maximum decoy velocity component is set to be the largest value that can be reliably tracked given the velocity uncertainty, i.e., $\vref=\vmax - \beta^v = 39$m/s. The resulting estimated positioning times are greater than the time it takes for the threats to reach the target jamming locations, $\hat{T}_{i^*_l,j^*_l} > \Tv_{j^*_l}$, for all assigned decoy-threat pairs, $(i^*_l,j^*_l)$, $l=\{1,\dots,n\}$. \cref{assum:ViableAssignment} therefore holds.

\definecolor{decoy1}{rgb}{0.2000,         0,    0.9000}%
\definecolor{decoy2}{rgb}{0.2857,    0.5357,    0.7857}%
\definecolor{decoy3}{rgb}{0.3714,         0,    0.6714}%
\definecolor{decoy4}{rgb}{0.4571,    0.5071,    0.5571}%
\definecolor{decoy5}{rgb}{0.5429,         0,    0.4429}%
\definecolor{decoy6}{rgb}{0.6286,    0.4786,    0.3286}%
\definecolor{decoy7}{rgb}{0.7143,         0,    0.2143}%
\definecolor{decoy8}{rgb}{0.8000,    0.4500,    0.1000}%
  \begin{table}%[th]
  \begin{center}
  {\footnotesize
  \begin{tabular}{>{\raggedleft\arraybackslash}p{0.7cm}|>{\raggedleft\arraybackslash}p{0.7cm}|>{\raggedleft\arraybackslash}p{1.9cm}|>{\raggedleft\arraybackslash}p{2.1cm}|>{\raggedleft\arraybackslash}p{2.0cm}|>{\raggedleft\arraybackslash}p{3.1cm}|>{\raggedleft\arraybackslash}p{3.1cm}}
%   {r|r|r|r|r|r}
  %begin{tabular}{c|c|c|c|c}
    Decoy & Threat & Bottleneck~order & Distance~to~target & Bound~saturation & Estimated~positioning~time & Optimised~positioning~time\\
    $i^*_l$ & $j^*_l$ & $l$  & $\|g_{j^*_l}-p_{i^*_l}(0)\|_\infty$ & $A_{i^*_l}$ & $\hat{T}_{i^*_l,j^*_l}$ & $\hat{N}^{\positioning*}_{i^*_l,j^*_l}[0] T_s$\\
    % ($i^*_k$ if assigned) &  $j^*_k$ &  $k$ &  $\mu_k$  & $T_{i^*_k,j^*_k}$  & $A_k$ ($A_8$ if unassigned) %$o_k$\\
    % \\%& $w_{i^c_k,j^c_k} - w_{i^*_k,j^*_k}$\\ 
    \hline
    \rowcolor{decoy1!50}
    $\delta_1$ & $\tau_4$  & 5 & 856m & 1555m & 24.0s & 12s \\
    \rowcolor{decoy2!50}
    $\delta_2$ & $\tau_1$ & 4 & 1227m & 1816m & 33.5s & 24s \\
    \rowcolor{decoy3!50}
    $\delta_3$ & $\tau_2$ & 6 & 589m & 1555m & 17.1s & 10s \\
    \rowcolor{decoy4!50}
    $\delta_4$ & none & none & no target & 1555m & & \\
     \rowcolor{decoy5!50}
    $\delta_5$ & $\tau_3$  & 1 & 1748m & 2249m & 46.8s & 36s \\
    \rowcolor{decoy6!50}
    $\delta_6$ & $\tau_6$ & 2 & 1566m & 1986m & 42.2s & 32s \\
    \rowcolor{decoy7!50}
    $\delta_7$ & none  & none & no target & 1555m & & \\
    \rowcolor{decoy8!50}
    $\delta_8$ & $\tau_5$ & 3 & 1439m & 1986m & 38.7s & 24s \\
  \end{tabular}
  }
  \end{center}
   \caption{Allocation of threats $\threats$ to decoys $\decoys$ based on sequential bottleneck assignment with weights, $\Delta_{i,j} = \|g_{j}-p_{i}(0)\|_\infty$, $(i,j)\in\decoys\times\threats $, with initial decoys positions, $p_i(0)$, target jamming locations, $g_j$, and colour coding shown in~\cref{fig:Initial}.}
       \label{tab:assignment}
\end{table} 

\begin{figure}%[th]
    \centering
        \begin{subfigure}{0.53715481171\linewidth} %width of pdf 1 / (width of pdf 1 + width of pdf 2) * 0.98\linewidth = 131 / (131 + 108) * 0.98\linewidth
      \centering
      \includegraphics[width=\linewidth]{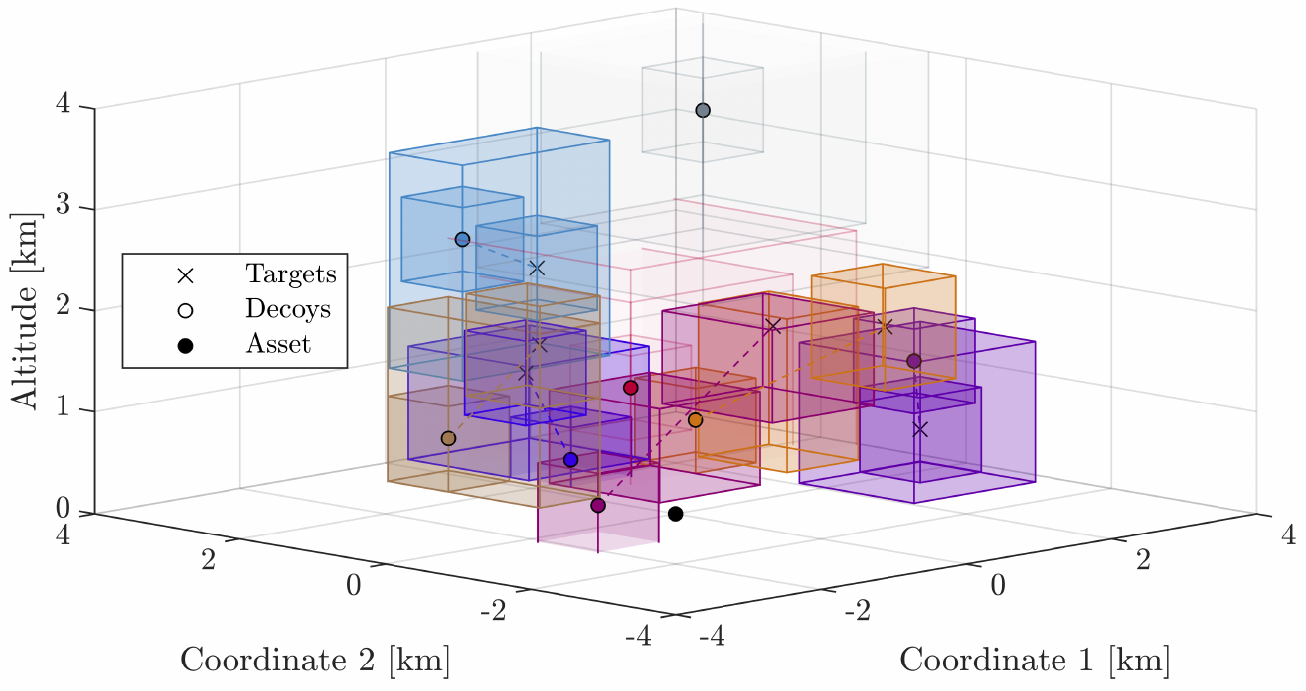}
      \caption{Side view} 
        \label{fig:SafeSets_SideView}
    \end{subfigure}
        \begin{subfigure}{0.44284518828\linewidth} %width of pdf 2 / (width of pdf 1 + width of pdf 2) * 0.98\linewidth = 108 / (131 + 108) * 0.98\linewidth
      \centering
      \includegraphics[width=\linewidth]{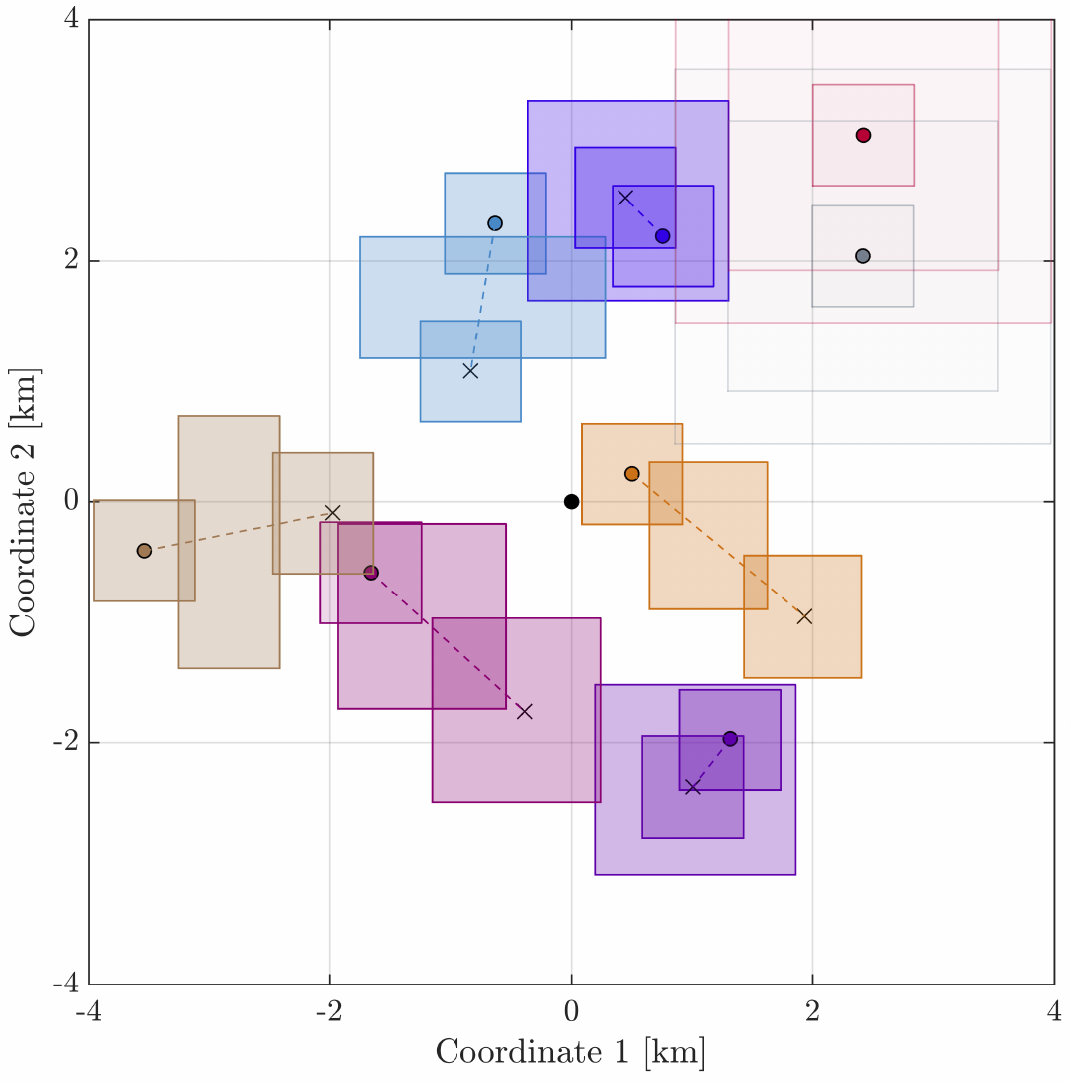}
      \caption{Top view} 
        \label{fig:SafeSets_TopView}
    \end{subfigure}
      \caption{Time-varying safe sets at $t=0$s, $t=20$s, and $t=40$s for initial decoy positions and target jamming locations shown in \cref{fig:Initial}.}
    \label{fig:SafeSets}
\end{figure}

 The decoy with the largest estimated positioning time is $i^*_1 = \delta_5$ assigned to threat $j^*_1=\tau_3$. Decoys $\delta_4$ and $\delta_7$ are not assigned to any threat. The smallest robustness margin is greater than the required safety distance between decoys, $\mumin = 838\textrm{m} > d$, satisfying \cref{assum:asslexglobalsafedist}. It follows that collision can be avoided by imposing the local position constraints specified in \cref{sec:Assignment}. The time-varying safe sets, $\locCon_{i}(t,d)$, for all decoys, $i \in \decoys$, are illustrated in \cref{fig:SafeSets}. We note that for any assigned decoy $i^*_l$, $l \in \{1,\dots,n\}$, the corresponding safe set, $\locCon_{i^*_l}(t,d)$, is larger when the there is a greater difference between the bound saturation value and the assignment weight, i.e., when $A_{i^*_l} - \dist_{i^*_l,j^*_l}$ is increased.
  
\subsection{Motion Planning}
The local safe sets allow for a decoupling of the motion control. For each decoy, a separate motion planning problem is solved. Decoys $\delta_4$ and $\delta_7$ are unassigned and are only required to remain within the local position constraint sets, $\locCon_{\delta_4}(t,d)$ and $\locCon_{\delta_7}(t,d)$ respectively, to not interfere with the positioning of the other decoys. That can be achieved by keeping $\delta_4$ and $\delta_7$ stationary. Given the estimated first-order bottleneck positioning time, $T_{\delta_5,\tau_3}=46.8$s, a horizon of $N=24$ discrete steps, corresponding to a planning period of $T=48$s, is chosen. We note that for some decoys the estimated time to complete the positioning task is significantly smaller. However, by finding an input sequence for which the decoys stay within the individual safe sets, a guarantee for the existence of collision-free motion control is obtained. 

For each assigned decoy, $i^*_l$, $l \in \{1,\dots,n\}$, the planning problem $\widehat{\operatorname{\ac{mptp}}}_{i^*_l,j^*_l}[0]$ is solved to obtain a sequence of commanded velocity inputs, ${\uv^*}_{i^*_l}^{N-1}[0]$, that minimises the predicted positioning time, $\hat{N}^{\positioning*}_{i^*_l,j^*_l}[0]T_s$. Following the encoding of the problem derived in \cref{sec:RobOpt}, optimisation tools Yalmip~\cite{Lofberg2004Yalmip} and Gurobi~\cite{gurobi} are used to solve the mixed-integer optimisation problems each with 200 binary auxiliary variables. For each decoy, an optimal solution is found in less than 1s on an Intel i7 CPU running at 2.8Hz. We note that the assignment procedure used to decouple the motion control problems takes less than $0.02$s to execute on the same machine. However, the complexity of the assignment problem scales with the number of threats as described in~\cref{rem:AssignmentComplexity}. The complexity of the motion planning is independent of the number of threats but sensitive to the number of time steps considered in the time horizon.
  
The resulting optimal values of the times required for decoys to robustly satisfy the positioning tasks while remaining in the safe sets are given in \cref{tab:assignment}. The optimised worst-case positioning time is $\hat{N}^{\positioning*}_{\delta_5,\tau_3}[0]T_s = 36$s. We observe that for all assigned decoys the optimised positioning times are significantly smaller than the estimated times. This demonstrates the benefit of formulating the positioning problem as a temporal-logic combination of subtasks consisting of reaching time-varying sets which outperforms the simpler strategy of moving with maximal possible velocity to a static point.  
  
\subsection{Open-Loop Simulation}  
We first consider a simulation where all elements of the optimised input sequences of length $N$, i.e., ${\uv^*}_{i^*_l}^{N-1}[0] $, are applied, for all assigned decoys $i^*_l$, $l \in \{1,\dots,n\}$, while neglecting the motion of the asset according to \cref{assum:StraightThreat}. The decoy trajectories are realised by simulating predictive low-level controllers that run at 25Hz for a higher fidelity model of the decoy and satisfy \cref{assum:IntersampleBahaviour}. 

\begin{figure}%[th]
      \centering
      \begin{subfigure}{0.49\linewidth}
        \centering
         \includegraphics[width=\linewidth]{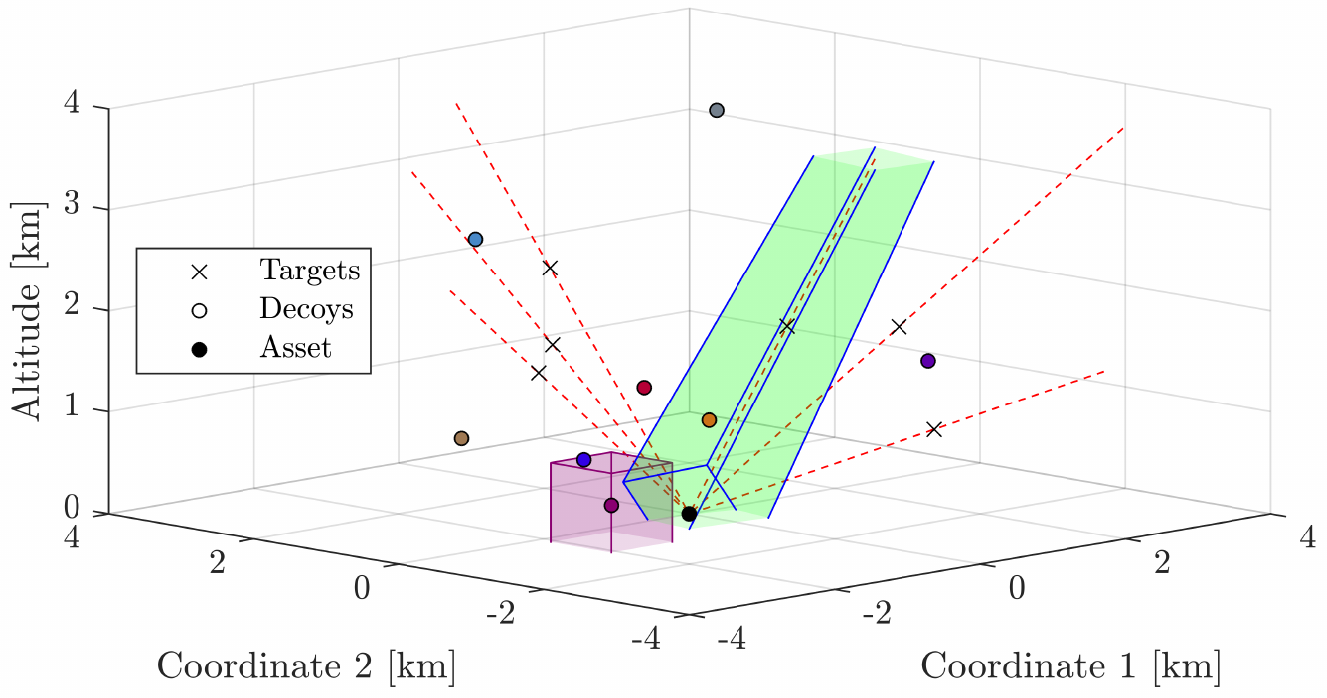}
        \caption{Initial condition at time step $k=0$ ($t=0$s)} 
        \label{fig:decoy5_k0}
    \end{subfigure} %\hfill
    \begin{subfigure}{0.49\linewidth}
        \centering
         \includegraphics[width=\linewidth]{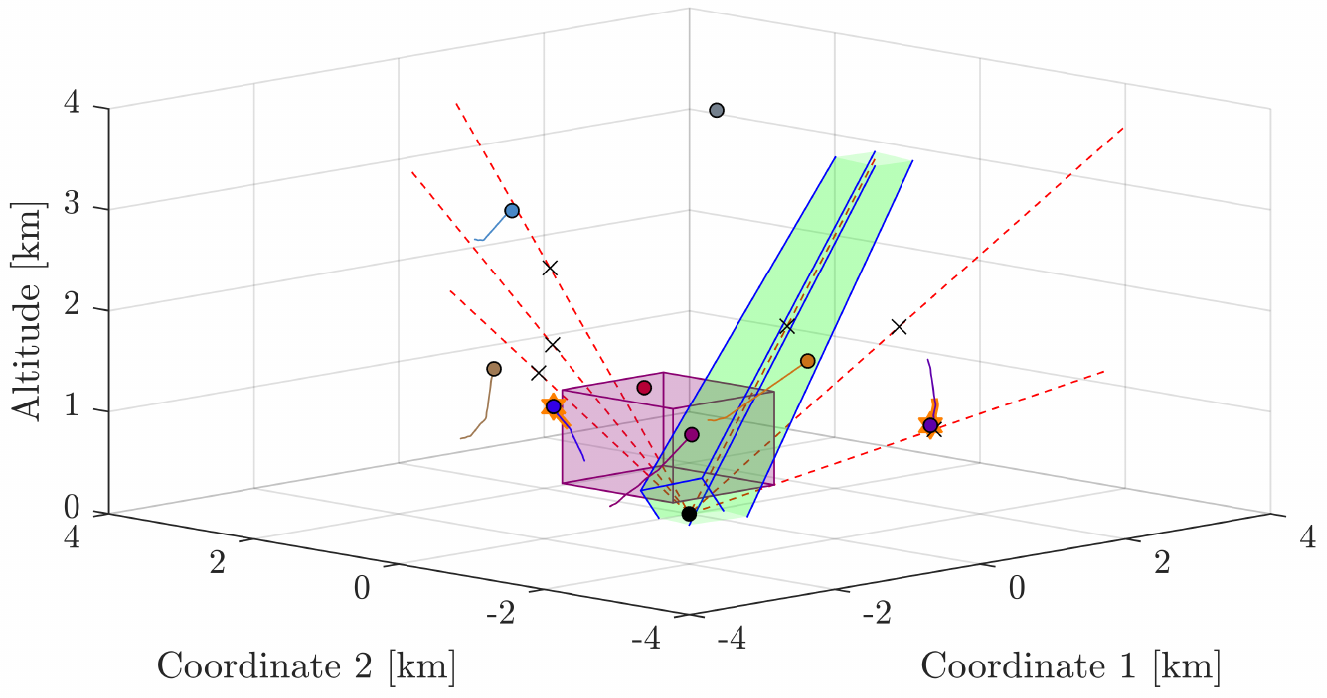}
        \caption{Condition at time step $k=10$ ($t = 20$s) } 
        \label{fig:decoys5_k10}
    \end{subfigure}
    \begin{subfigure}{0.49\linewidth}
        \centering
        \includegraphics[width=\linewidth]{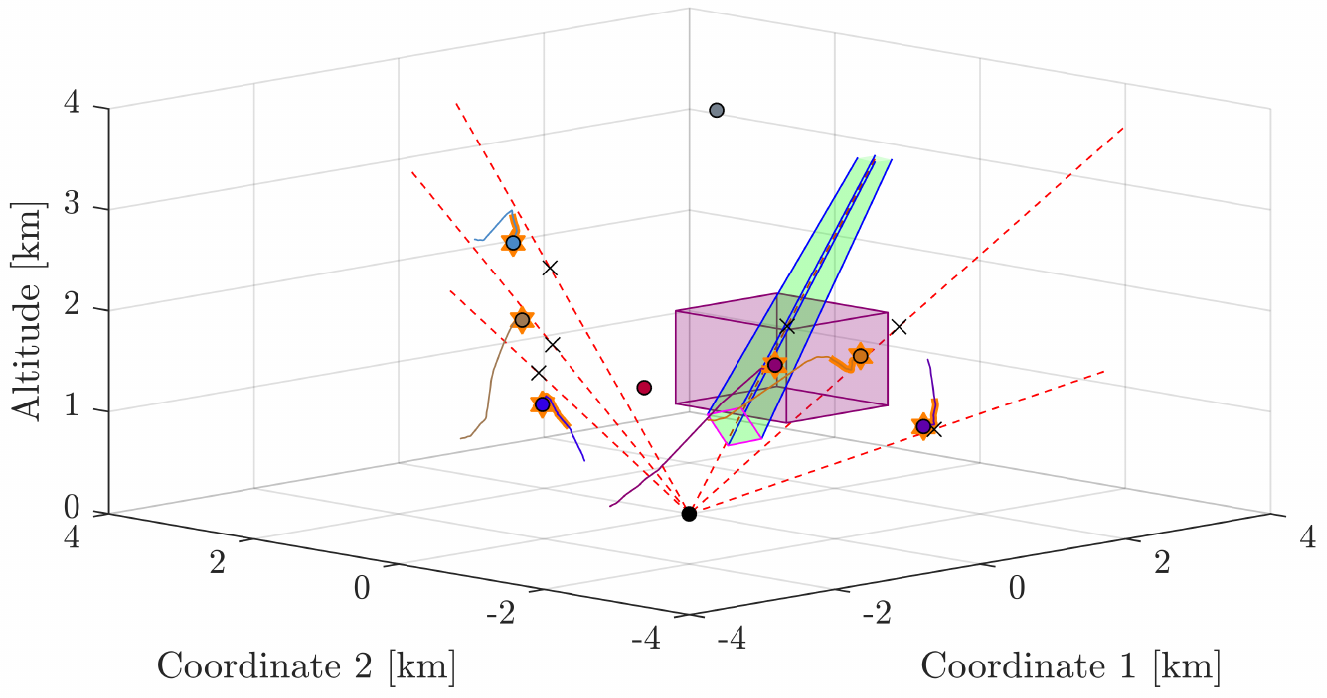}
        \caption{Condition at time step $k=20$ ($t = 40$s) } 
        \label{fig:decoy5_k20}
    \end{subfigure} %\hfill
    \begin{subfigure}{0.49\linewidth}
        \centering
        \includegraphics[width=\linewidth]{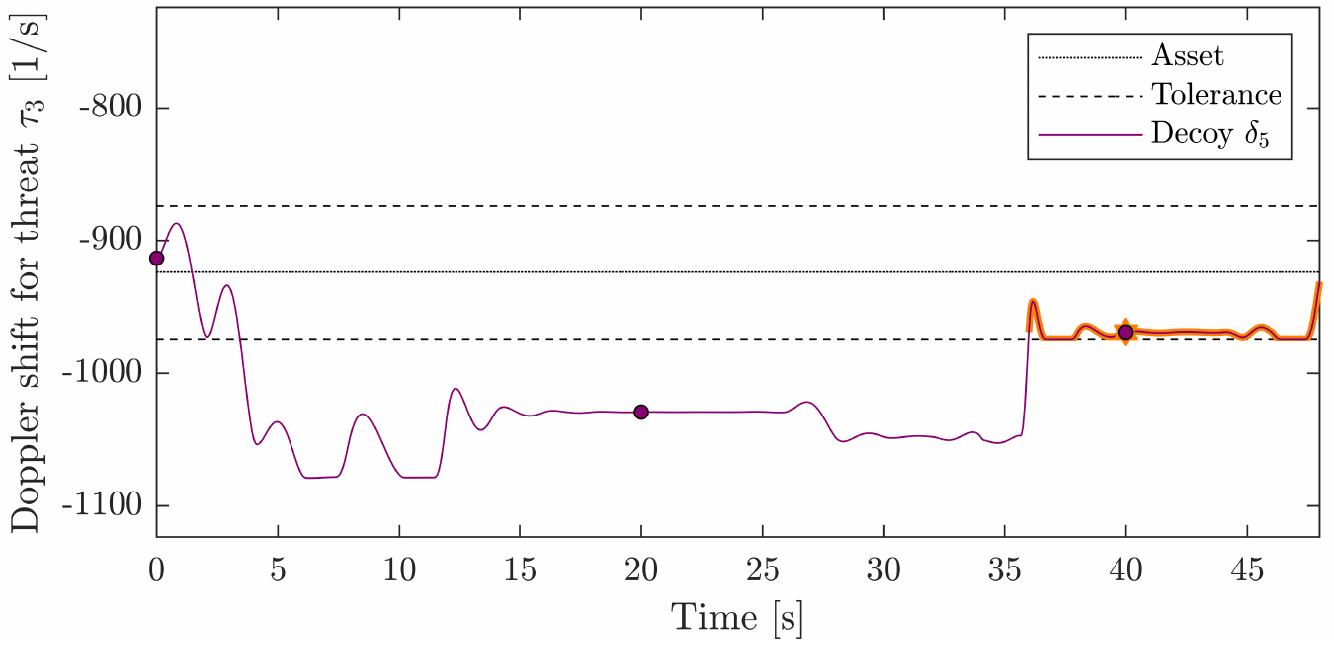}
        \caption{Comparison of Doppler shifts observed by threat $\tau_3$ caused by the asset and decoy $\delta_5$ over planning horizon %($t \in [0s,48s]$)
        } 
        \label{fig:DopplerShift_Threat3}
    \end{subfigure}
    \caption{Open-loop simulation of optimised positioning strategies with focus on decoy $\delta_5$ assigned to $\tau_3$ by illustrating the safe set, $\locCon_{\delta_5}(t,d)$ with decoy-coloured surface, and the region suitable for jamming, $\ApproxTrackingCone_{\tau_3}(t) \setminus \ApproxBurnThroughSet_{\tau_3}(t)$ with green surface (see \cref{fig:TrackingCone_hitset}); trajectories %of decoys 
    are highlighted with orange when positioning has been completed.}
      \label{fig:openloop_decoy5}
  \end{figure}

\cref{fig:openloop_decoy5} shows the resulting decoy trajectories with a particular focus on the first-order bottleneck decoy, $\delta_5$, for which the safe set, $\locCon_{\delta_5}(t,d)$, and the inner-approximation of the region in which jamming of threat $\tau_3$ can be initialised, $\ApproxTrackingCone_{\tau_3}(t) \setminus \ApproxBurnThroughSet_{\tau_3}(t)$, are illustrated. The variation of the potential Doppler shift $\delta_5$ would cause for $\tau_3$ is shown in \cref{fig:DopplerShift_Threat3}. We observe that after time step $k=18$ the Doppler shift caused by $\delta_5$ remains within the bound for it to be indistinguishable from the asset, i.e., $|\DopplerShift_{\delta_5,\tau_3}(t)-\DopplerShift_{\asset,\tau_3}| \leq \maxdsvar$ for all $t \in [36\textrm{s},48\textrm{s}]$. Because in addition $p_{\delta_5}(36\textrm{s}) \in \TrackingCone_{\tau_3}(36\textrm{s}) \setminus \BurnThroughSet_{\tau_3}(36\textrm{s})$, the simulation confirms that seduction of $\tau_3$ can be initialised at $t=36$s. Furthermore, since all decoys $i \in \decoys$, are positioned inside the corresponding safe sets, $p_i(t) \in \locCon_i(t,d)$, throughout all of the planning horizon $t \in [0,48\textrm{s}]$, and the unassigned decoys $\delta_4,\delta_7$ remain stationary, none of the decoys collide with each other.

\subsection{Closed-Loop Simulation}  
Given that $\widehat{\operatorname{\ac{mptp}}}_{i^*_l,j^*_l}[0]$ can be solved individually for all assigned decoy-threat pairs, $(i^*_l,j^*_l)$, $l \in \{1,\dots,n\}$, in less than the sampling time, $T_s$, the motion planning can be repeated for every decoy in every discrete time step with updated information on the states of the decoy, threat, and asset to deal with unaccounted uncertainties and model mismatches. In a closed-loop implementation, the planning problem is applied with a shrinking horizon, i.e., at each time step $k \in \{0,1,\dots,N-2\}$, the planning problem $\widehat{\operatorname{\ac{mptp}}}_{i^*_l,j^*_l}[k]$ is solved for each $l \in \{1,\dots,n\}$ in parallel to obtain a sequence of commanded velocity inputs of length $N-k$ of which only the first element, ${u^*}_{i^*_l}[k]$, is applied before solving $\widehat{\operatorname{\ac{mptp}}}_{i^*_l,j^*_l}[k + 1]$, with a reduced prediction horizon of $N-1-k$, to obtain an updated value of  ${u^*}_{i^*_l}[k + 1]$ in the next time step. In this way, feedback information on the trajectories of the threats and the asset and on the realisations of the disturbances in the decoy model is incorporated. 

We consider a closed-loop simulation in which the asset is moving with a constant velocity of $v^\asset=
  \begin{bmatrix}
     0 & -5 &  0
  \end{bmatrix}^\top$\textrm{m/s} along the ground surface. While the motion planning assumes the asset to be stationary, the simulated asset position changes over time. As a consequence, the simulated threat trajectories also deviate from the assumed straight-line trajectories. Specifically, a threat is simulated to initially regulate its velocity vector to follow the time-varying direction towards the asset, as defined in \eqref{eq:ThreatVelocity_pre}. At the time when a decoy starts jamming the threat, the threat heading control switches to tracking the direction towards the moving fake asset portrayed by the decoy. In this simulation, the control strategy for each decoy is switched as soon as the decoy completes the positioning task. To lure the assigned threat away from the asset, the decoy is controlled such that its velocity component orthogonal to the threat velocity is maximised while assuring that the Doppler shift observed by the threat matches the one caused by the asset.

\begin{figure}%[h]
    \centering
        \begin{subfigure}{0.53269709543\linewidth} %width of pdf 1 / (width of pdf 1 + width of pdf 2) * 0.98\linewidth = 131 / (131 + 110) * 0.98\linewidth
      \centering
      \includegraphics[width=\linewidth]{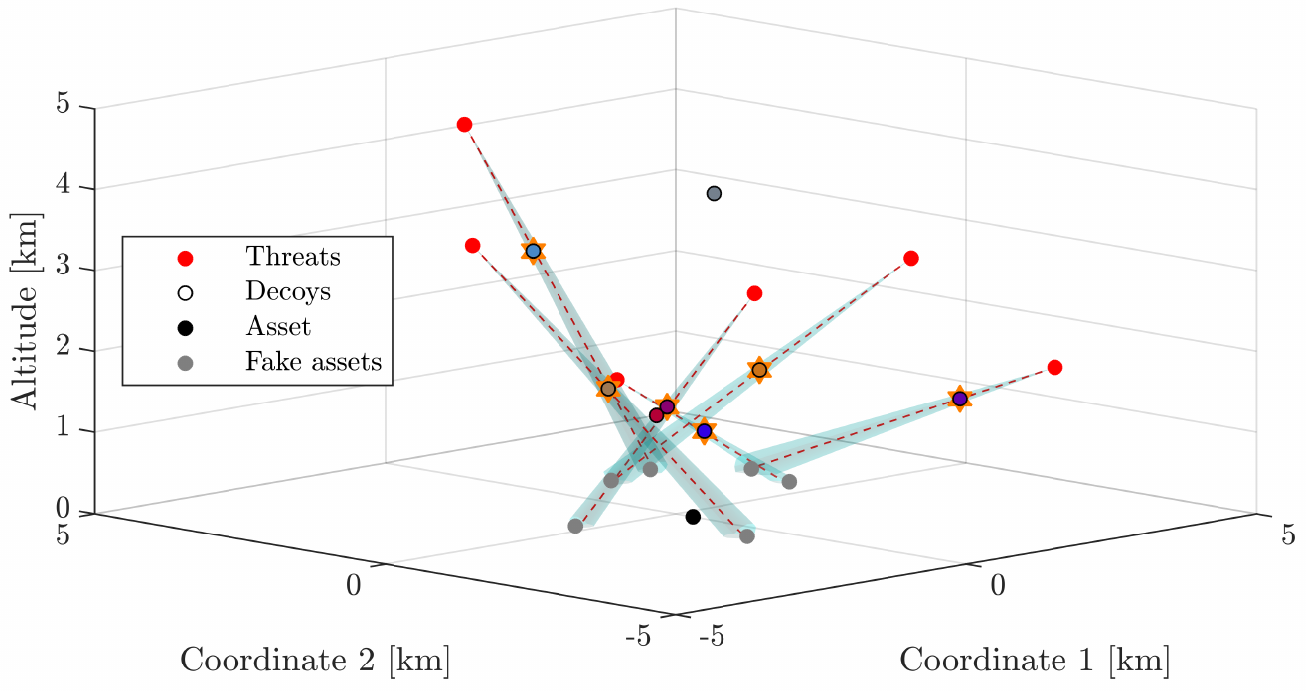}
      \caption{Configuration of threats, asset, decoys and the resulting fake assets at time $t = 60$s.} 
        \label{fig:ClosedLoopk30}
    \end{subfigure}
        \begin{subfigure}{0.44730290456\linewidth} %width of pdf 2 / (width of pdf 1 + width of pdf 2) * 0.98\linewidth = 110 / (131 + 110) * 0.98\linewidth
      \centering
      \includegraphics[width=\linewidth]{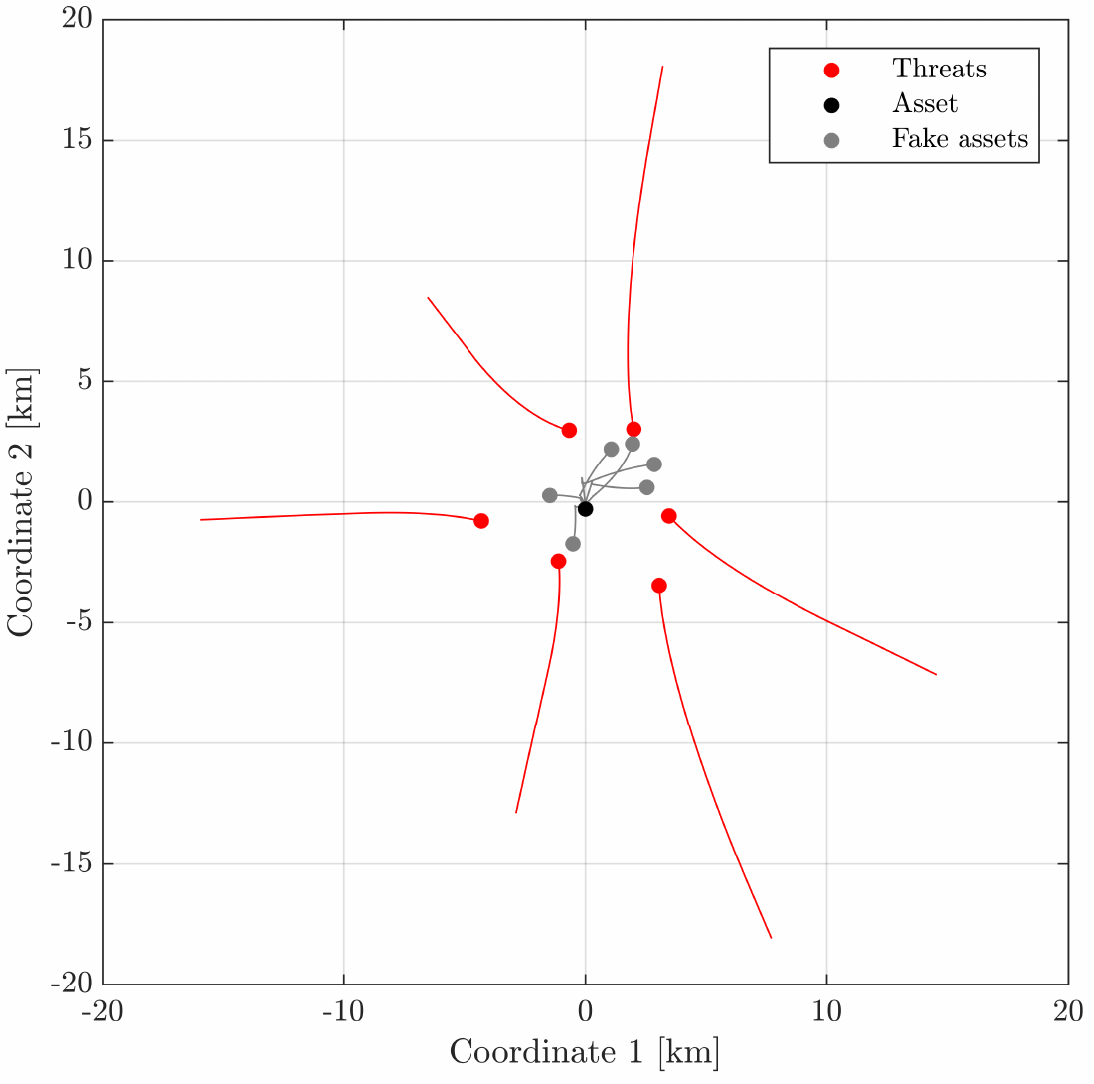}
      \caption{Top view of threat and fake asset trajectories} 
        \label{fig:SeductionTrajectoryk30}
    \end{subfigure}
      \caption{Result of closed-loop simulation for the initial configuration given in \cref{fig:Initial}.}
    \label{fig:ClosedLoop}
\end{figure}

 \cref{fig:ClosedLoop} shows the configuration of the threats, decoys, and asset at time $t=60$s in a closed-loop simulation with the same initial condition shown in \cref{fig:Initial} and assignment given in \cref{tab:assignment}. The simulation reveals that by providing feedback on the environment to repeatedly update the motion planning, some deviations from the planning assumptions can be tolerated. In particular, in the case of small deviations of the asset and threat trajectories, specifically ones of comparable size to the uncertainty considered in the decoy model, the shrinking horizon implementation results in the same decoy positioning times as obtained in the open-loop analysis. Furthermore, we observe that after 60s the decoys have successfully lured all threats away from the asset as can be seen by the trajectories of the portrayed fake assets shown in \cref{fig:SeductionTrajectoryk30}.       
 
\section{Conclusions}\label{sec:conclusion}
We derived a strategy to position multiple threat-seduction decoys in minimum time. The conditions required for a decoy to deceive a threat were modelled as time-varying set-membership constraints on the decoy position and velocity. In particular, the three conditions of the decoy being in the tracking cone, the threat being outside of the burn-through range, and the Doppler shift from the decoy being indistinguishable from the one from the asset were considered. 

To decide which decoys deals with which threat, we applied an assignment approach that lexicographically minimises the estimated time it takes for decoys to travel to suitable fixed locations where the threats can be deceived. We derived a closed-form solution for determining such locations that stay suitable for jamming for the longest possible duration. The assignment procedure provides a quantification of the amount of deviation from straight-line constant-speed trajectories that can be tolerated. Given that this robustness margin is sufficiently large in relation to the size of the decoys, local affine position constraints are obtained that guarantee the avoidance of inter-agent collisions. We exploited the local safe sets to decouple the motion control of the individual decoys. 

To be able to address the complex requirements of the motion control, we used a crude for the decoy dynamics but accounted for its inaccuracy by including robustness to model uncertainty in the planning. We formulated a minimum-time positioning problem for an individual decoy with the sequence of commanded velocities at discrete time steps over a finite planning horizon as optimisation variables. The positioning task is modelled as an \ac{ltl} specification where the conditions required for successfully jamming a threat are each represented by an atomic proposition and combined with temporal logic. For each atomic proposition, we derived associated time-varying affine constraints on the decoy state. To combine the atomic propositions we coupled the different state constraints by introducing auxiliary mixed-integer variables. With these models, we encoded the robust minimum-time planning problem as an \ac{milp} that can be solved online. In a simulated case study we demonstrated the ability of the method to minimise the time required to deceptively jam multiple threats simultaneously. The result showcases the benefit of having temporal logic constraints that do not predetermine when the jamming conditions have to be satisfied and incorporate this flexibility into the optimisation instead. 

The focus of this work was on positioning multiple decoys as fast as possible. The proposed method provides a certificate in scenarios where the assignment is viable, sufficiently robust, and the planning problem is feasible. In cases where no such certificate is provided, there is the potential to exploit partial components of the method but further research is required to incorporate alternative approaches when some components are not guaranteed to be successful, e.g., alternative collision avoidance approaches when the safe sets are too limiting. Optimising the placing of the decoys at the initial time is the topic of ongoing work. Another open problem of interest is optimal control and coordination of decoys in the seduction phase after positioning has successfully been completed.

\appendices
\section{Proof of \cref{thm:TargetLocation}}\label{app:ProofTargetLocation}
% \begin{proof}
We first show that if a position, $p\in\R^3$, satisfies the tracking cone and the no-burn-through conditions at time $t$, then the projection of $p$ onto the line through the asset position and the threat position, $p' = z_j(t) + \frac{(p-z_j(t))^\top(y(t)-z_j(t))}{\|y(t)-z_j(t)\|_2^2}\left(y(t)-z_j(t)\right)$, also satisfies these conditions at time $t$. To this end, we assume $p\in\TrackingCone_j(t)$. From \cref{assum:StraightThreat} the threat velocity is $\dot{z}_j(t) = \frac{\SpeedThreat}{\|y(t)-z_j(t)\|_2} \left(y(t)-z_j(t)\right)$ and therefore we have  $\left(p'-z_j(t)\right)^\top \dot{z}_j(t)= \left(p-z_j(t)\right)^\top \dot{z}_j(t)$. From \eqref{eq:trackingcone} we know that $\left(p-z_j(t)\right)^\top \dot{z}_j(t)\leq \|p-z_j(t)\|_2\SpeedThreat\cos(\theta)$. It follows that $\|p'-z_j(t)\|_2 \leq \|p-z_j(t)\|_2 \cos(\theta)$. Because $0<\theta<\frac{\pi}{2}$, we have $\|p'-z_j(t)\|_2\leq \|p-z_j(t)\|_2$ and conclude that $p'\in\TrackingCone_j(t)$. Furthermore, assuming that the no-burn-through condition is satisfied for decoy position $p$ at time $t$, i.e., $\|y(t)-z_j(t)\|\geq\elconst_j\sqrt{\|p-z_j(t)\|_2}$, it follows that $\|y(t)-z_j(t)\|_2\geq\elconst_j\sqrt{\|p'-z_j(t)\|_2}$ and thus that there is also no burn through for decoy position $p'$ at time $t$. 

It is therefore sufficient to only consider points on the line $l_j:=\{q\in\R^3\,|\,\exists\upsilon\in\R: q = \upsilon z_j(0) +  (1 - \upsilon)y(0)\}$ in order to find a location that stays viable for starting the seduction process for the largest possible time period. We parametrise an optimising point on $l_j$ as $g_j = \upsilon^*_{j} z_j(0) + (1 - \upsilon^*_{j}) y(0)$. From \cref{assum:StraightThreat} we know that $z_j(t)=z_j(0) + \frac{\SpeedThreat t}{\|y(0)-z_j(0)\|_2}\left(y(0)-z_j(0)\right)$. For $g_j$ to lie in the tracking cone at time $t=0$ the line parameter has to be bounded by $0\leq\upsilon^*_{j}\leq 1$. For a given value of $\upsilon^*_{j}\in[0,1]$, the location $g_j$ remains in the tracking cone for a duration of $\Tc_j(\upsilon^*_{j}) = \frac{(1-\upsilon^*_{j})\|y(0)-z_j(0)\|_2}{\SpeedThreat}$, i.e we have $g_j\in\TrackingCone_j(t)$ for all $t\in[0,\Tc_j(\upsilon^*_{j})]$.

We can express the satisfaction of the no-burn-through condition for a decoy at position $g_j$ at time $t$ as $\|y(0)-z_j(0)\|_2 - \SpeedThreat t \geq \elconst_j\sqrt{(1-\upsilon^*_{j})\|y(0)-z_j(0)\|_2 -\SpeedThreat t}$. We define $Q_{j}(t,\upsilon):={\SpeedThreat}^2\left(t - \frac{2\|y(0)-z_j(0)\|_2-{\elconst_j}^2}{2\SpeedThreat}\right)^2 - \frac{\elconst^4}{4}+\upsilon\elconst^2\|y(0)-z_j(0)\|_2$ and equivalently rewrite the condition for no burn-through occurring as $Q_{j}(t,\upsilon^*_j)\geq 0$. For any $\upsilon$, the function $Q_{j}(t,\upsilon)$ is minimised with respect to the time parameter $t$ at value $\tb_{j} :=\frac{2\|y(0)-z_j(0)\|_2-\elconst_j^2}{2\SpeedThreat}$. We see that $Q_{j}(\tb_{j},\upsilon)\geq 0$ if $\upsilon\geq \lb_{j} := \frac{\elconst_j^2}{4\|y(0)-z_j(0)\|_2}$. Thus, for $\upsilon^*_{j} = \lb_{j}$ burn through never occurs. It follows that for decoy position $g_j=y(0) - \frac{\elconst_j^2}{4\|y(0)-z_j(0)\|_2}\left(y(0)-z_j(0)\right)$ both the tracking cone and the no-burn-through conditions are satisfied for a duration of ${\Tv_{j}} = \Tc_{j}(\lb_{j})$, i.e., $g_j\in\TrackingCone_j(t)$ and $Q_{j}(t)\geq 0$ for all $t\in[0,{\Tv_{j}}]$. 

It remains to show that there is no value for the parameter $\upsilon^*_{j}$ for which the corresponding location $g_j$ is viable to break the lock of the threat for a longer duration than for $\upsilon^*_{j}=\lb_{j}$. If $\upsilon^*_{j}> \lb_{j}$, we see that $\Tc(\upsilon^*_{j})<\Tc(\lb_{j})$ and thus that the tracking cone condition is satisfied for a shorter duration than for $\lb_{j}$. We now assume for the sake of contraction that there exists a $\upsilon'$, with $\upsilon'<\lb_{j}$, for which $\upsilon'z_j(0)+(1-\upsilon')y(0)\in\TrackingCone_j(t)$ and $Q(t,\upsilon')\geq 0$ for all $t\in[0,t']$, where $t' > \Tc_{j}(\lb_{j})$. For $\upsilon'<\lb_{j}$ burn through occurs at time $\tb_{j}$ because $Q(\tb,\upsilon')<Q(\tb,\lb)=0$. Considering that $\Tc_j(\lb_{j}) = \frac{4\|y(0)-z_j(0)\|_2-\elconst_j^2}{4\SpeedThreat}\geq \tb_{j} = \frac{4\|y(0)-z_j(0)\|_2-2\elconst_j^2}{4\SpeedThreat}$, it is evident that there is a contradiction and that there does not exit a target decoy position that remains suited for initialising the seduction of threat $j$ for longer than $g_j$ given in~\eqref{eq:targetpoint}. 
% \end{proof} 
\hfill $\qed$

\section{Constraints to Encode Positioning Specification}\label{sec:MIconstraints}
Considering a decoy $i \in \decoys$ with state sequence $\xv_i^{N}[0]$, we derive the constraints required to encode the positioning specification $\varphi_j^\positioning$ for the seduction of threat~$j \in \threats$. The atomic propositions, ${ap}_j^\cone$, ${ap}_j^\bt$, and ${ap}_j^\doppler$ are defined in~\cref{sec:TrackingCone,,sec:BurnThrough,,sec:DopplerShift}, respectively. Given a bounded set containing all considered values of the decoy state, $\bar{\StateConstraints} \subset \R^6$, the polyhedral sets associated with the atomic propositions can be expressed as
$\polytope_j^\cone[k] = \big\{ x \in \bar{\StateConstraints} \,\big|\, \Phi_{{ap}_j^\cone}[k] x \leq \Psi_{{ap}_j^\cone}[k] \big\}$, 
with $\Phi_{{ap}_j^\cone}[k] \in \R^{n_\cone \times 6}$, $\Psi_{{ap}_j^\cone}[k] \in \R^{n_\cone }$, 
where $n_\cone = 5$, 
$\polytope_j^\bt[k] = \big\{ x \in \bar{\StateConstraints} \,\big|\, \Phi_{{ap}_j^\bt}[k] x \leq \Psi_{{ap}_j^\bt}[k] \big\}$, 
with $\Phi_{{ap}_j^\bt}[k] \in \R^{n_\bt \times 6}$, $\Psi_{{ap}_j^\bt}[k] \in \R^{n_\bt }$,
where $n_\bt = 1$, 
and $\polytope_j^\doppler[k] = \big\{ x \in \bar{\StateConstraints} \,\big|\, \Phi_{{ap}_j^\doppler}[k] x \leq \Psi_{{ap}_j^\doppler}[k] \big\}$, with $\Phi_{{ap}_j^\doppler}[k] \in \R^{n_\doppler \times 6}$, $\Psi_{{ap}_j^\doppler}[k] \in \R^{n_\doppler }$, where $n_\doppler = 2$, 
for all $k \in \{0,1,\dots,N\}$. 
We apply the Big-M reformulation~\cite{Bemporad1999Automatica} to encode atomic propositions. 

For ${ap}_j^\cone$ and ${ap}_j^\doppler$, we introduce binary vectors, 
$\beta_{{ap}_j^\cone}[k] \in \{0,1\}^{n_\cone}$, $\beta_{{ap}_j^\doppler}[k] \in \{0,1\}^{n_\doppler}$, and impose 
\begin{align}
\Phi_{{ap}_j^\cone}[k] x_i[k] &\leq \Psi_{{ap}_j^\cone}[k] + M_{{ap}_j^\cone}^+\big(\1_{n_\cone} -\beta_{{ap}_j^\cone}[k]\big)\,,\\
\Phi_{{ap}_j^\doppler}[k] x_i[k] &\leq \Psi_{{ap}_j^\doppler}[k] + M_{{ap}_j^\doppler}^+\big(\1_{n_\doppler} -\beta_{{ap}_j^\doppler}[k]\big)\,,
\end{align} 
for all time steps $k \in \{0,1,\dots,N\}$, with $M_{{ap}_j^\cone}^+ > \max_{x \in \bar{\StateConstraints}} \max_{k\in\{0,1,\dots,N\}} \max_{q \in \{1,\dots,n_\cone\} }  \big[\Phi_{{ap}_j^\cone}[k] x - \Psi_{{ap}_j^\cone}[k]\big]_q$ and $M_{{ap}_j^\doppler}^+ > \max_{x \in \bar{\StateConstraints}} \max_{k\in\{0,1,\dots,N\}} \max_{q \in \{1,\dots,n_\doppler\} }  \big[ \Phi_{{ap}_j^\doppler}[k] x - \Psi_{{ap}_j^\doppler}[k] \big]_q$, where $\1_{n_a}$ is a vector of size $n_a$ with all elements equal to 1. For all $k\in\{0,1,\dots,N\}$, we then introduce bounded continuous scalar variables $0 \leq \gamma_{{ap}_j^\cone}[k] \leq 1$, $0 \leq \gamma_{{ap}_j^\doppler}[k] \leq 1$, and constrain
\begin{align}
    \gamma_{{ap}_j^\cone}[k] \1_{n_\cone} & \leq \beta_{{ap}_j^\cone}[k]\,,
    \\ \gamma_{{ap}_j^\doppler}[k] \1_{n_\doppler} & \leq \beta_{{ap}_j^\doppler}[k]\,,
    \\ \gamma_{{ap}_j^\cone}[k] & \geq 1 - n_\cone + \1_{n_\cone}^\top \beta_{{ap}_j^\cone}[k]\,,
    \\ \gamma_{{ap}_j^\doppler}[k] & \geq 1 - n_\doppler + \1_{n_\doppler}^\top  \beta_{{ap}_j^\doppler}[k]\,.
\end{align}  
As a result, the continuous variables, $\gamma_{{ap}_j^\cone}[k]$, $\gamma_{{ap}_j^\doppler}[k]$, only take values in the binary set $\{0,1\}$, $\gamma_{{ap}_j^\cone}[k] = 1$ implies $\xv_i^{N-k}[k] \satisfies {ap}_j^\cone$, and $\gamma_{{ap}_j^\doppler}[k] = 1$ implies $\xv_i^{N-k}[k] \satisfies {ap}_j^\doppler$ for all $k\in\{0,1,\dots,N\}$.

To encode ${ap}_j^\bt$, we introduce one binary $\beta_{{ap}_j^\bt}[k]$ for every time step $k\in\{0,1,\dots,N\}$. Because only the negation of ${{ap}_j^\bt}$ is relevant for the satisfaction of $\varphi_j^\positioning$, we impose
\begin{align}
    \Phi_{{ap}_j^\bt}[k] x_i[k] &> \Psi_{{ap}_j^\bt}[k] + M_{{ap}_j^\bt}^- \beta_{{ap}_j^\cone}[k] \,,
\end{align} for all $k\in\{0,1,\dots,N\}$, with $M_{{ap}_j^\bt}^- < \min_{x \in \bar{\StateConstraints}} \min_{k\in\{0,1,\dots,N\}} \min_{q \in \{1,\dots,n_\cone\} }  \big[\Phi_{{ap}_j^\bt}[k] x - \Psi_{{ap}_j^\bt}[k]\big]_q$.  For all $k\in\{0,1,\dots,N\}$, we then have that $\beta_{{ap}_j^\bt}[k] = 0$ implies $ \xv_i^{N-k}[k] \satisfies \lnot{ap}_j^\bt$.

To encode the formula $\lalways {ap}^\doppler_j$, we introduce bounded continuous variables, $0 \leq \gamma_{\lalways{ap}^\doppler_j}[k] \leq 1$, for all $k\in\{0,1,\dots,N\}$. 
By setting $\gamma_{\lalways{ap}^\doppler_j}[N] = \gamma_{{ap}^\doppler_j}[N]$ and for $k\in\{0,1,\dots,N-1\}$ constraining
\begin{subequations}\label{eq:gammaalwaysdoppler}
\begin{align+}
     \gamma_{\lalways{ap}^\doppler_j}[k] &\leq \gamma_{{ap}^\doppler_j}[k]\,,
     \\ \gamma_{\lalways{ap}^\doppler_j}[k] &\leq \gamma_{\lalways{ap}^\doppler_j}[k+1]\,,
     \\ \gamma_{\lalways{ap}^\doppler_j}[k] &\geq -1 + \gamma_{{ap}^\doppler_j}[k] +  \gamma_{\lalways{ap}^\doppler_j}[k+1]\,,
\end{align+}
\end{subequations}
we enforce $\gamma_{\lalways{ap}^\doppler_j}[k]\in\{0,1\}$ and that $\gamma_{\lalways{ap}^\doppler_j}[k] = 1$ implies $\xv_i^{N-k}[k] \satisfies \lalways {ap}^\doppler_j$.

 To the encode the formula $\bar{\varphi}_j={ap}^\cone_j \land \lnot{ap}^\bt_j \land \lalways{ap}^\doppler_j$, we introduce additional continuous variables $0 \leq \gamma_{\bar{\varphi}_j}[k] \leq 1$ for all $k\in\{0,1,\dots,N\}$.  
By imposing the following constraints,
\begin{align}
    \gamma_{\bar{\varphi}_j}[k] &\leq \gamma_{{ap}^\cone_j}[k]\,,
    \\ \gamma_{\bar{\varphi}_j}[k] &\leq 1 - \beta_{{ap}^\bt_j}[k]\,,
    \\ \gamma_{\bar{\varphi}_j}[k] &\leq \gamma_{\lalways{ap}^\doppler_j}[k]\,,
    \\ \gamma_{\bar{\varphi}_j}[k] &\geq -1 + \gamma_{{ap}^\cone_j}[k] - \beta_{{ap}^\bt_j}[k] + \gamma_{\lalways{ap}^\doppler_j}[k]\,,
\end{align} for all $k\in\{0,1,\dots,N\}$, we have $\gamma_{\bar{\varphi}_j}[k]\in\{0,1\}$ and that $\gamma_{\bar{\varphi}_j}[k] = 1$ implies $\xv_i^{N-k}[k] \satisfies \bar{\varphi}_j$. 
% \begin{align}
%     \gamma_{\bar{\varphi}_j}[k] &= 1 & \Rightarrow & &  \xv_i^{N-k}[k] &\satisfies \bar{\varphi}_j\,.
% \end{align} 

Finally, to encode the specification $\varphi^\positioning_{j}=\leventually\bar{\varphi}_j$, we introduce the continuous variables $0 \leq \gamma_{\varphi^\positioning_{j}}[k] \leq 1$ for $k\in\{0,1,\dots,N\}$. We set $\gamma_{\varphi^\positioning_{j}}[k] = \gamma_{\bar{\varphi}_j}[N]$ and constrain
% \begin{subequations}\label{eq:eventually_encoding}
\begin{align}
    \gamma_{\varphi^\positioning_{j}}[k] &\geq \gamma_{\bar{\varphi}_j}[k]\,,
    \\ \gamma_{\varphi^\positioning_{j}}[k] &\geq \gamma_{\bar{\varphi}_j}[k+1]\,,
    \\ \gamma_{\varphi^\positioning_{j}}[k] &\leq \gamma_{\bar{\varphi}_j}[k] + \gamma_{\bar{\varphi}_j}[k+1]\,,
\end{align}
% \end{subequations}
for all $k\in\{0,1,\dots,N-1\}$ to enforce $\gamma_{\varphi^\positioning_{j}}[k]\in\{0,1\}$ and that $\gamma_{\varphi_j^\positioning}[k] = 1$ implies $\xv_i^{N-k}[k] \satisfies \varphi_j^\positioning$.
% \begin{align}\label{eq:onewayencoding} 
%     \gamma_{\varphi_j^\positioning}[k] &= 1 & \Rightarrow & &  \xv_i^{N-k}[k] &\satisfies \varphi_j^\positioning\,.
% \end{align}

% \section*{Funding Sources}

\section*{Acknowledgments}
The authors would like to thank Daniel Gibbons, Anna Dostovalova, and Jijoong Kim for their valuable input and helpful discussions. This work was supported by the Defence Science and Technology Group through research agreements MyIP:7558, MyIP:7562, and MyIP:9156.

\bibliographystyle{IEEEtran}
\bibliography{FamilyPaper}

\begin{thebibliography}{10}
\providecommand{\url}[1]{#1}
\csname url@samestyle\endcsname
\providecommand{\newblock}{\relax}
\providecommand{\bibinfo}[2]{#2}
\providecommand{\BIBentrySTDinterwordspacing}{\spaceskip=0pt\relax}
\providecommand{\BIBentryALTinterwordstretchfactor}{4}
\providecommand{\BIBentryALTinterwordspacing}{\spaceskip=\fontdimen2\font plus
\BIBentryALTinterwordstretchfactor\fontdimen3\font minus
  \fontdimen4\font\relax}
\providecommand{\BIBforeignlanguage}[2]{{%
\expandafter\ifx\csname l@#1\endcsname\relax
\typeout{** WARNING: IEEEtran.bst: No hyphenation pattern has been}%
\typeout{** loaded for the language `#1'. Using the pattern for}%
\typeout{** the default language instead.}%
\else
\language=\csname l@#1\endcsname
\fi
#2}}
\providecommand{\BIBdecl}{\relax}
\BIBdecl

\bibitem{Adamy2000Book}
D.~L. Adamy, \emph{{EW 101: A First Course in Electronic Warfare}}.\hskip 1em
  plus 0.5em minus 0.4em\relax Artech House, 2001.

\bibitem{Shames2017CDC}
I.~Shames, A.~Dostovalova, J.~Kim, and H.~Hmam, ``{Task Allocation and Motion
  Control for Threat-Seduction Decoys},'' in \emph{Conference on Decision and
  Control}, 2017, pp. 4509--4514.

\bibitem{Shima2009Book}
T.~Shima and S.~Rasmussen, \emph{{UAV Cooperative Decision and Control:
  Challenges and Practical Approaches}}.\hskip 1em plus 0.5em minus 0.4em\relax
  SIAM, 2009.

\bibitem{Jang2019JoAE}
I.~Jang, H.~S. Shin, A.~Tsourdos, J.~Jeong, S.~Kim, and J.~Suk, ``{An
  Integrated Decision-Making Framework of a Heterogeneous Aerial Robotic Swarm
  for Cooperative Tasks with Minimum Requirements},'' \emph{Proceedings of the
  Institution of Mechanical Engineers, Part G: Journal of Aerospace
  Engineering}, vol. 233, no.~6, pp. 2101--2118, 2019.

\bibitem{Burkard2012Book}
R.~E. Burkard, M.~Dell'Amico, and S.~Martello, \emph{{Assignment Problems,
  revised reprint}}.\hskip 1em plus 0.5em minus 0.4em\relax Siam, 2012.

\bibitem{Pentico2007EJoOR}
D.~W. Pentico, ``{Assignment Problems: A Golden Anniversary Survey},''
  \emph{European Journal of Operational Research}, vol. 176, no.~2, pp.
  774--793, 2007.

\bibitem{Bertuccelli2009GNaCC}
L.~Bertuccelli, H.-L. Choi, P.~Cho, and J.~How, ``{Real-Time Multi-UAV Task
  Assignment in Dynamic and Uncertain Environments},'' in \emph{AIAA guidance,
  navigation, and control conference}, 2009, p. 5776.

\bibitem{Bethke2008RaAM}
B.~Bethke, M.~Valenti, and J.~P. How, ``{UAV Task Assignment},'' \emph{IEEE
  Robotics and Automation Magazine}, vol.~15, no.~1, 2008.

\bibitem{Edison2011CaOR}
E.~Edison and T.~Shima, ``{Integrated Task Assignment and Path Optimization for
  Cooperating Uninhabited Aerial Vehicles Using Genetic Algorithms},''
  \emph{Computers and Operations Research}, vol.~38, no.~1, pp. 340--356, 2011.

\bibitem{Morgan2016IJoRR}
D.~Morgan, G.~P. Subramanian, S.~J. Chung, and F.~Y. Hadaegh, ``{Swarm
  Assignment and Trajectory Optimization using Variable-Awarm, Distributed
  Auction Assignment and Sequential Convex Programming},'' \emph{International
  Journal of Robotics Research}, vol.~35, no.~10, pp. 1261--1285, 2016.

\bibitem{Turpin2014AR}
M.~Turpin, K.~Mohta, N.~Michael, and V.~Kumar, ``{Goal Assignment and
  Trajectory Planning for Large Teams of Interchangeable Robots},''
  \emph{Autonomous Robots}, vol.~37, no.~4, pp. 401--415, 2014.

\bibitem{Gravell2021CEP}
B.~Gravell and T.~Summers, ``{Centralized Collision-Free Polynomial
  Trajectories and Goal Assignment for Aerial Swarms},'' \emph{Control
  Engineering Practice}, vol. 109, no. February, p. 104753, 2021.

\bibitem{Bertrand2014AJ}
S.~Bertrand, J.~Marzat, A.~Kahn, and Y.~Rochefort, ``{MPC Strategies for
  Cooperative Guidance of Autonomous Vehicles},'' \emph{AerospaceLab journal},
  vol.~8, pp. 1--18, 2014.

\bibitem{Wu2019ISMRMAS}
F.~Wu, V.~S. Varadharajan, and G.~Beltrame, ``{Collision-Aware Task Assignment
  for Multi-Robot Systems},'' in \emph{International Symposium on Multi-Robot
  and Multi-Agent Systems}, 2019, pp. 30--36.

\bibitem{Shames2011IJoRaNC}
I.~Shames, B.~Fidan, and B.~D. Anderson, ``{Close Target Reconnaissance with
  Guaranteed Collision Avoidance},'' \emph{International Journal of Robust and
  Nonlinear Control}, vol.~21, no.~16, pp. 1823--1840, 2011.

\bibitem{MacAlpine2015CoAI}
P.~MacAlpine, E.~Price, and P.~Stone, ``{SCRAM: Scalable Collision-Avoiding
  Role Assignment with Minimal-makespan for formational positioning},'' in
  \emph{Proceedings of the National Conference on Artificial Intelligence},
  vol.~3, 2015, pp. 2096--2102.

\bibitem{Turpin2014IJRR}
M.~Turpin, N.~Michael, and V.~Kumar, ``{CAPT: Concurrent Assignment and
  Planning of Trajectories for Multiple Robots},'' \emph{International Journal
  of Robotics Research}, vol.~33, no.~1, pp. 98--112, 2014.

\bibitem{Wood2020RAL}
T.~A. Wood, M.~Khoo, E.~Michael, C.~Manzie, and I.~Shames, ``{Collision
  Avoidance Based on Robust Lexicographic Task Assignment},'' \emph{IEEE
  Robotics and Automation Letters}, vol.~5, no.~4, pp. 5693--5700, 2020.

\bibitem{Tabuada2005TAC}
P.~Tabuada and G.~J. Pappas, ``{Linear Time Logic Control of Discrete-Time
  Linear Systems},'' \emph{IEEE Transactions on Automatic Control}, vol.~51,
  no.~12, pp. 1862--1877, 2006.

\bibitem{Belta2007RaAM}
C.~Belta, A.~Bicchi, M.~Egerstedt, E.~Frazzoli, E.~Klavins, and G.~J. Pappas,
  ``{Symbolic Planning and Control of Robot Motion [Grand Challenges of
  Robotics]},'' \emph{IEEE Robotics and Automation Magazine}, vol.~14, no.~1,
  pp. 61--70, 2007.

\bibitem{Kress2009ToR}
H.~Kress-Gazit, G.~E. Fainekos, and G.~J. Pappas, ``{Temporal-Logic-Based
  Reactive Mission and Motion Planning},'' \emph{IEEE Transactions on
  Robotics}, vol.~25, no.~6, pp. 1370--1381, 2009.

\bibitem{Wongpiromsarn2012TAC}
T.~Wongpiromsarn, U.~Topcu, and R.~M. Murray, ``{Receding Horizon Temporal
  Logic Planning},'' \emph{IEEE Transactions on Automatic Control}, vol.~57,
  no.~11, pp. 2817--2830, 2012.

\bibitem{Karaman2008CDC}
S.~Karaman, R.~G. Sanfelice, and E.~Frazzoli, ``{Optimal Control of Mixed
  Logical Dynamical Systems with Linear Temporal Logic Specifications},''
  \emph{Conference on Decision and Control}, pp. 2117--2122, 2008.

\bibitem{Wolff2014ICRA}
E.~M. Wolff, U.~Topcu, and R.~M. Murray, ``{Optimization-Based Trajectory
  Generation with Linear Temporal Logic Specifications},'' in
  \emph{International Conference on Robotics and Automation}, 2014, pp.
  5319--5325.

\bibitem{Gol2015Automatica}
E.~{Aydin Gol}, M.~Lazar, and C.~Belta, ``{Temporal Logic Model Predictive
  Control},'' \emph{Automatica}, vol.~56, pp. 78--85, 2015.

\bibitem{Frick2017LCSS}
D.~Frick, T.~A. Wood, G.~Ulli, and M.~Kamgarpour, ``{Specifications in
  Uncertain Environments},'' \emph{IEEE Control Systems Letters}, vol.~1,
  no.~1, pp. 20--25, 2017.

\bibitem{Sessa2018HSCC}
P.~G. Sessa, D.~Frick, T.~A. Wood, and M.~Kamgarpour, ``{From Uncertainty Data
  to Robust Policies for Temporal Logic Planning},'' in \emph{Conference on
  Hybrid Systems: Computation and Control}, 2018, pp. 157--166.

\bibitem{Wolff2012CDC}
E.~M. Wolff, U.~Topcu, and R.~M. Murray, ``{Robust Control of Uncertain Markov
  Decision Processes with Temporal Logic Specifications},'' \emph{Conference on
  Decision and Control}, pp. 3372--3379, 2012.

\bibitem{Bloem2012JCSS}
R.~Bloem, B.~Jobstmann, N.~Piterman, A.~Pnueli, and Y.~Sa'Ar, ``{Synthesis of
  Reactive(1) Designs},'' \emph{Journal of Computer and System Sciences},
  vol.~78, no.~3, pp. 911--938, 2012.

\bibitem{Ding2011IFAC}
D.~{Ding, Xu Chu Dennis and Smith, Stephen L. and Belta, Calin and Rus}, ``{LTL
  Control in Uncertain Environments with Probabilistic Satisfaction
  Guarantees},'' in \emph{IFAC World Congress}, vol.~44, 2011, pp. 3515--3520.

\bibitem{Lahijanian2011ToR}
M.~Lahijanian, S.~B. Andersson, and C.~Belta, ``{Probabilistic Satisfaction
  Guarantees},'' \emph{IEEE Transactions on Robotics}, vol.~28, no.~2, pp.
  1--14, 2011.

\bibitem{Kamgarpour2017Automatica}
M.~Kamgarpour, T.~A. Wood, S.~Summers, and J.~Lygeros, ``{Control Synthesis for
  Stochastic Systems Given Automata Specifications Defined by Stochastic
  Sets},'' \emph{Automatica}, vol.~76, pp. 177--182, 2017.

\bibitem{Michael2021arXiv}
E.~Michael, T.~A. Wood, C.~Manzie, and I.~Shames, ``{Global Sensitivity
  Analysis for Bottleneck Assignment Problems},'' \emph{arXiv preprint
  arXiv:2104.00803}, 2021.

\bibitem{Baier2008book}
C.~Baier and J.-P. Katoen, \emph{{Principles of Model Checking}}.\hskip 1em
  plus 0.5em minus 0.4em\relax MIT press, 2008.

\bibitem{Birkedal2006Arix}
L.~Birkedal, R.~E. M{\o}gelberg, and R.~{Lerchedahl Petersen C}, ``{Linear
  Abadi and Plotkin Logic},'' \emph{Logical Methods in Computer Science},
  vol.~2, no.~2, pp. 1--48, 2006.

\bibitem{Kaptan2012Thesis}
A.~Kaptan, ``{Net-centric controlled distributed stand-in-jamming using
  UAVS-Transmission losses and range limitations due to geo-localization
  problem ver turkish geography},'' Ph.D. dissertation, Naval Postgraduate
  School, Monterey, California, 2012.

\bibitem{Shekhar2015}
R.~C. Shekhar, M.~Kearney, and I.~Shames, ``{Robust Model Predictive Control of
  Unmanned Aerial Vehicles Using Waysets},'' \emph{Journal of Guidance,
  Control, and Dynamics}, vol.~38, no.~10, pp. 1898--1907, 2015.

\bibitem{Khoo2020arXiv}
M.~Khoo, T.~A. Wood, C.~Manzie, and I.~Shames, ``{A Greedy and Distributable
  Approach to the Lexicographical Bottleneck Assignment Problem with Conditions
  on Exactness},'' \emph{arXiv preprint arXiv:2008.12508}, 2020.

\bibitem{Bertsimas2011SIAMr}
D.~Bertsimas, D.~B. Brown, and C.~Caramanis, ``{Theory and applications of
  robust optimization},'' \emph{SIAM review}, vol.~53, no.~3, pp. 464--501,
  2011.

\bibitem{Lofberg2004Yalmip}
J.~L{\"{o}}fberg, ``{YALMIP : A Toolbox for Modeling and Optimization in
  MATLAB},'' in \emph{In Proceedings of the CACSD Conference}, Taipei, Taiwan,
  2004.

\bibitem{gurobi}
\BIBentryALTinterwordspacing
{Optimization Gurobi, LCC}, ``{Gurobi Optimizer Reference Manual},'' Tech.
  Rep., 2021. [Online]. Available: \url{http://www.gurobi.com}
\BIBentrySTDinterwordspacing

\bibitem{Bemporad1999Automatica}
A.~Bemporad and M.~Morari, ``{Control of Systems Integrating Logic, Dynamics,
  and Constraints},'' \emph{Automatica}, vol.~35, no.~3, pp. 407--427, 1999.

\end{thebibliography}

\end{document}